\documentclass[twocolumn,superscriptaddress,
 amsmath,amssymb,aps,prr,floatfix]{revtex4-2}
\usepackage[usenames,dvipsnames]{color}
\newcommand{\aop}{\hat a}
\newcommand{\adop}{\hat a ^\dagger}
\usepackage{comment}
\usepackage{cancel}
\usepackage{ulem}
\usepackage{physics}
\usepackage{graphicx}
\usepackage{dcolumn}
\usepackage{bm}
\usepackage[breaklinks=true,colorlinks,citecolor=blue,linkcolor=blue,urlcolor=blue]{hyperref}
\usepackage[percent]{overpic}
\usepackage{tensor}
\begin{document}
\title{Photon condensation, Van Vleck paramagnetism, and chiral cavities}
\author{Alberto Mercurio}
\thanks{These three authors contributed equally.}
\affiliation{Dipartimento di Scienze Matematiche e Informatiche, Scienze Fisiche e  Scienze della Terra, 
Universit\`{a} di Messina, I-98166 Messina,~Italy}
\author{Gian Marcello Andolina}
\thanks{These three authors contributed equally.}
\affiliation{ICFO-Institut de Ci\`{e}ncies Fot\`{o}niques, The Barcelona Institute of Science and Technology, Av. Carl Friedrich Gauss 3, 08860 Castelldefels (Barcelona),~Spain}
\author{Francesco M. D. Pellegrino}
\thanks{These three authors contributed equally.}
\affiliation{Dipartimento di Fisica e Astronomia ``Ettore Majorana'', Universit\`a di Catania, Via S. Sofia 64, I-95123 Catania,~Italy}
\affiliation{INFN, Sez.~Catania, I-95123 Catania,~Italy}
\affiliation{CNR-IMM, Via S. Sofia 64, I-95123 Catania, Italy}
\author{Omar Di Stefano}
\affiliation{Dipartimento di Scienze Matematiche e Informatiche, Scienze Fisiche e  Scienze della Terra, Universit\`{a} di Messina, I-98166 Messina,~Italy}
\author{Pablo Jarillo-Herrero}
\affiliation{Department of Physics, Massachusetts Institute of Technology, Cambridge, Massachusetts,~USA}
\author{Claudia Felser}
\affiliation{Max Planck Institute for Chemical Physics of Solids, N\"{o}thnitzer Str. 40, Dresden 01187,~Germany}
\author{Frank H. L. Koppens}
\affiliation{ICFO-Institut de Ci\`{e}ncies Fot\`{o}niques, The Barcelona Institute of Science and Technology, Av. Carl Friedrich Gauss 3, 08860 Castelldefels (Barcelona),~Spain}
\affiliation{ICREA-Instituci\'{o} Catalana de Recerca i Estudis Avan\c{c}ats, Passeig de Llu\'{i}s Companys 23, 08010 Barcelona,~Spain}
\author{Salvatore Savasta}
\affiliation{Dipartimento di Scienze Matematiche e Informatiche, Scienze Fisiche e  Scienze della Terra, 
Universit\`{a} di Messina, I-98166 Messina,~Italy}
\author{Marco Polini}
\affiliation{Dipartimento di Fisica dell'Universit\`a di Pisa, Largo Bruno Pontecorvo 3, I-56127 Pisa,~Italy}
\affiliation{Istituto Italiano di Tecnologia, Graphene Labs, Via Morego 30, I-16163 Genova,~Italy}
\affiliation{ICFO-Institut de Ci\`{e}ncies Fot\`{o}niques, The Barcelona Institute of Science and Technology, Av. Carl Friedrich Gauss 3, 08860 Castelldefels (Barcelona),~Spain}
\begin{abstract}
We introduce a gauge-invariant model of planar, square molecules coupled to a quantized spatially-varying cavity electromagnetic vector potential $\hat{\bm A}({\bm r})$. Specifically, we choose a temporally {\it chiral} cavity hosting a uniform magnetic field $\hat{\bm B}$, as this is 
the simplest instance in which a transverse spatially-varying $\hat{\bm A}({\bm r})$ is at play. We show that when the molecules are in the Van Vleck paramagnetic regime, an equilibrium quantum phase transition to a photon condensate state occurs. 
\end{abstract}
\maketitle

\section{Introduction}

Basic quantum statistical mechanics dictates that free massless photons, despite being bosonic particles, cannot condense into a single macroscopic state~\cite{Huang}. On the other hand, when photons are confined to a cavity~\cite{Raimond01,Haroche13,cavityqm,BlochReview,PCavities} and coupled to matter degrees of freedom (such as excitons), condensation into a single quantum state can occur~\cite{carusotto_rmp_2013}. In this Letter, we are interested in the ground state of a photon condensate~\cite{footnote_Dicke,hepp_lieb,wang_pra_1973}, i.e.~a state containing a macroscopically large number of coherent 
photons, {\it i.e.}~$\langle\hat{a}\rangle\propto \sqrt{N}$, where $\hat{a}$ ($\hat{a}^\dagger$) destroys (creates) a cavity photon. This phase transition is forbidden by gauge invariance when the vector potential $\hat{\bm A}$ describing the electromagnetic properties of the cavity is spatially uniform~\cite{rzazewski_prl_1975,rzazewski_prl_2006,tufarelli_pra_2015,nataf_naturecommun_2010,viehmann_prl_2011,chirolli_prl_2012,pellegrino_prb_2014,Rubio2018,DeBernardis18,DeBernardis19,mazza_prl_2019,Schafer,andolina_prb_2019,andolina_epjp_2022}.
\begin{figure}[t]
\centering
\begin{overpic}[width= \linewidth]{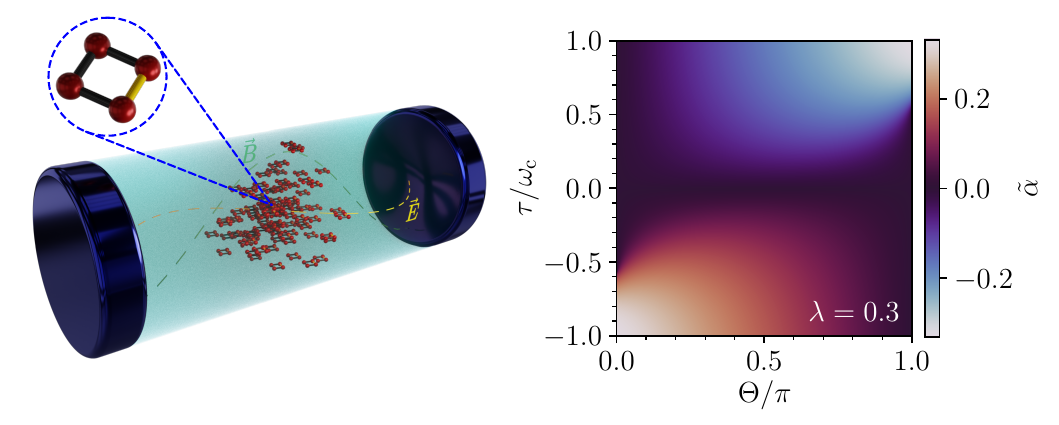}
\put(-2,37){\footnotesize{(a)}}\put(45,37){\footnotesize{(b)}}
\end{overpic}
\caption{(Color online) (a)  Pictorial representation of the setup  where many planar square molecules interact with a cavity's single-mode, quantized magnetic field, $\hat{\bm B}$. The field is perpendicular to the molecular plane. Due to the small spatial extension of the molecule cloud compared to the field's wavelength, the magnetic field is assumed constant and the electric field negligible.
Note that, in each molecule, one of the hopping integrals (yellow) is different from the other three ones. (b) Phase diagram of the model. Results in this figure have been obtained by setting with $t = \omega_{\rm c}$ and $\lambda=0.3$. The photon condensate order parameter $\tilde{\alpha}$ (color scale) is plotted as a function of the two microscopic parameters $\tau \in  [-t,t]$ and $\Theta \in [0, \pi[$. \label{fig: 3d_fig}
}
\end{figure}

Recent studies, however, have shown that when {\it itinerant} electron systems are coupled in a gauge-invariant fashion to a transverse {\it spatially-varying} electromagnetic vector potential $\hat{\bm A}({\bm r})$, equilibrium photon condensation may occur as a magnetostatic instability~\cite{nataf_prl_2019,andolina_prb_2020,guerci_prl_2020,Basko22,Mazza23}. Key to this phenomenon is the orbital paramagnetic character of the electronic system. Orbital paramagnetism, i.e.~a positive sign of the orbital magnetic susceptibility, is a rather rare phenomenon in nature, as diamagnetism tends to dominate~\cite{Huang,mcclure_pr_1956}. Nevertheless, itinerant electron systems may display orbital paramagnetism~\cite{vignale_prl_1991,bruder_prl_1998,principi_prl_2010,gomez_prl_2011,raoux_prl_2014,piechon_prb_2016}.

On a seemingly disconnected path, the impact of vacuum cavity fields on the chemical and physical properties of molecules has been demonstrated~\cite{VacuumReview,ebbesen_acr_2016,kowalewski_pnas_2017,fregoni_acsnano_2022}. A great deal of interest emerged after seminal experiments unveiled how photochemical reaction rates can be modified within cavities~\cite{thomas_science_2019,hutchison_angewchem_2012}.
This blooming field is nowadays known as {\it polaritonic chemistry}~\cite{fregoni_acsnano_2022}.  Fundamental theoretical work in this field~\cite{haugland_jcp_2021,Haugland_PRX_2020,riso_nc_2022,ruggenthaler_arxiv_2021,rokaj_jpb_2018,schafer_acs_2020}  is carried out in the framework of the electrical dipole approximation, whereby molecular transitions couple solely with the cavity electric field, thereby neglecting magnetic effects. The need to transcend the electrical dipole approximation in polaritonic chemistry is beginning to emerge. For example, the authors of Ref.~\cite{riso_arxiv_2022} go beyond it in order to distinguish between two enantiomers in chiral cavities.

The key question we try and answer in this Letter is the following: can photon condensation---which, so far, has been studied only in itinerant electron systems~\cite{nataf_prl_2019,andolina_prb_2020,guerci_prl_2020}---occur in the realm of polaritonic chemistry? In this Letter, we show that incorporating the effects of magnetic coupling, photon condensation can occur in a system of molecules, which does not display extended Bloch states and an itinerant character. Two conditions need to be satisfied. On the one hand, one needs to work with a molecular system whose orbital response displays paramagnetic character. To this end, we exploit a genuinely quantum mechanical mechanism leading to orbital paramagnetism in a molecular system, which is often dubbed ``Van Vleck paramagnetism"~\cite{Grosso}. This mechanism guarantees a paramagnetic orbital response provided that the molecule has a doublet of quasi-degenerate levels. On the other hand, one needs to confine these Van Vleck molecules to a cavity where a transverse spatially-varying ${\bm A}({\bm r})$ is at play. Since a spatially-varying ${\bm A}$ leads to a finite magnetic field ${\bm B}({\bm r}) =\nabla_{\bm r} \times {\bm A}({\bm r})$, the simplest cavity one can consider is a cavity hosting a uniform ${\bm B}$ field. Our cavity falls into the category of chiral cavities~\cite{hubener_naturematerials_2021} with chirality of temporal (rather than structural) character.

\section{Model}

We consider a single, spinless electron~\cite{footnotespin} hopping between the $n_{\rm s}$ sites of a planar plaquette, such as a square or triangular plaquette, lying in the $\hat{\bm x}$-$\hat{\bm y}$ plane.(Here, $\hat{\bm x}$, $\hat{\bm y}$, and $\hat{\bm z}$ are unit vectors in the $x$, $y$, and $z$ directions, respectively.) This system will be considered as a toy model of a ``molecule" since the plaquette contains a small number of sites ($n_{\rm s} = 4$ below) and the electron roaming in the plaquette is therefore far away from the lattice regime where electronic Bloch bands emerge. 

We then consider a system of $N$ such molecules described by the following electronic Hamiltonian, $\hat{\mathcal{H}}_{\rm e} =   \sum^N_{k=1}  \hat{h}_{{\rm e},k}$ , where
\begin{equation}\label{eq:single_molecule_Hamiltonian_no_cavity}
\hat{h}_{{\rm e},k}=  - \sum_{j = 0}^{n_{\rm s}  -1} \left(  t_{j,j+1} \hat{c}_{j,k}^\dagger \hat{c}_{j+1,k}+ {\rm H.c.} \right)
\end{equation}
is the single-molecule Hamiltonian. Here, $\hat{c}_{j,k}^\dagger$ ($\hat{c}_{j,k}$) creates (destroys) an electron on the $j$-th site of the $k$-th plaquette, $t_{j,j+1}=\delta_{j,0} t^\prime+(1-\delta_{j,0})t$ is the tunnelling amplitude between neighboring sites of a plaquette, and 
\begin{equation}
t^\prime=\tau e^{-i \Theta}~,
\end{equation}
where $\tau \in {[-t,t]}$ (with $t{ >}0$ is assumed to be real, without any loss of generality) and $\Theta \in [0,\pi[$ are the tunneling ``intensity'' and phase, respectively. We emphasize that we have taken $t^\prime \neq t$ to simulate different physical conditions. For $\Theta \neq 0$ ($\Theta \in ]0,\pi[$, yielding therefore ${\rm Im}[t'] \neq 0$), the electronic Hamiltonian is not invariant under time-reversal symmetry (TRS), and the plaquette supports a circulating ground-state current. Viceversa, when $\Theta=0$ the system is invariant under TRS and no ground-state currents flow. As discussed in Appendix~\ref{subsec: sm external field}, Hamiltonians like the one in Eq.~(\ref{eq:single_molecule_Hamiltonian_no_cavity}), which break TRS, can be physically obtained by applying an external, classical magnetic field along the $\hat{\bm z}$ direction. Note that, counterintuitively, also the particular case $\Theta=0$ and $t'=-t$ (which respects TRS) can be obtained by applying a particular classical transverse static magnetic field ${\bm B}_{\rm cl}$ to a system of square molecules that, in the absence of such field, have four identical hopping parameters $t_{j,j+1}=t~\forall j$, as detailed in Appendix~\ref{subsec: sm external field}. Note that, despite the presence of ${\bm B}_{\rm cl}$, the Hamiltonian in this case respects TRS. This occurs for a special magnetic field, ${\bm B}_{\rm cl} = \pm B_{\pi} \hat{\bm z}$, with $B_{\pi}= c \pi / (e A_{\rm p})$. Consider, indeed, the case in which one is in the presence of ${\bm B}_{\rm cl} = \pm B_{\pi} \hat{\bm z}$. In both cases, it is easy to check that one gets $\Theta=0$ and $t'=-t$. Now, applying TRS has the net result of changing the classical field from ${\bm B}_{\rm cl}$ into $-{\bm B}_{\rm cl} = \mp B_{\pi} \hat{\bm z}$. We conclude that at the fields ${\bm B}_{\rm cl} = \pm B_{\pi} \hat{\bm z}$ one gets a Hamiltonian that is invariant under TRS.

We now couple the molecular system to a single-mode cavity with a non-vanishing magnetic field oriented along the $z$ direction. 
The dynamics of the total system is governed by the following Hamiltonian
\begin{align}
    \hat{\mathcal{H}} =   \omega_{\rm c} \adop \aop 
       { -} \sum_{k, j}\left( t_{j,j+1} e^{-i \theta_{j,j+1}}  \hat{c}_{j,k}^\dagger \hat{c}_{j+1,k}+ {\rm H.c.} \right)~,
\end{align}
where $\omega_{\rm c}$ is the cavity photon energy ($\hbar =1$), 
and the Peierls phase $\theta_{j, j + 1} = (-e/c) \int_{{\bm r}_j}^{{\bm r}_{j + 1}} {\bm A}({\bm r}) \cdot d^2 {\bm r}$  is necessary to minimally couple the matter degrees of freedom living on the plaquette to the cavity field~\cite{footnotespin}.  Here, for the sake of simplicity, we consider a system composed of planar plaquettes with all the same orientation, so that the tunneling coefficients are independent of the specific $k$ molecule. In principle, a system of molecules with the same orientation can be realized by growing a molecular crystal (see, e.g., Ref.~\cite{ghirri_arxiv_2302.00804}).

In the symmetric gauge, the vector potential of the single photon mode is ${\bm A}({\bm r}) = - B y \hat{\bm x} / 2 + B x \hat{\bm y} / 2$, with $\nabla_{\bm r} \times {\bm A}({\bm r}) = {\bm B} = B \hat{\bm z}$. Quantization of the cavity field is carried out in the usual manner by promoting $B$ from a $c$-number to a bosonic quantum operator $B \rightarrow B_0 (\aop + \adop)$, where $\hat{a}^\dagger$ ($\hat{a}$) creates (destroys) a cavity photon.
The final Hamiltonian describing light-matter interactions in the cavity is:

\begin{align}
\label{eq: initial hamiltonian}
    \hat{\mathcal{H}} = \omega_{\rm c} \adop \aop { -}  \sum_{k,j}\left[t_{j,j+1} e^{-i  \lambda  ( \aop + \adop)/{\sqrt{N}}} \hat{c}_{j,k}^\dagger \hat{c}_{j+1,k}+ {\rm H.c.} \right]~,\end{align}
where $\lambda=-  2\pi [\Phi/ (\Phi_0n_s)] \sqrt{N}$ is a dimensionless light-matter coupling constant, which is proportional to the ratio between the magnetic flux  $\Phi \equiv B_0 A_{\rm p}$ piercing the plaquette of area $A_{\rm p}$ and the flux quantum $\Phi_0 = 2 \pi c/e$. For example, for a square plaquette ($n_{\rm s} = 4$) of side $d$, $A_{\rm p} = d^2$. Note that i) in the thermodynamic $N\to \infty$ limit, $\lambda$ is independent of $N$, since $B_0$ scales as $\sim 1/ \sqrt{N}$ to make sure that the magnetic field $B=B_0 (\aop + \adop)$ is an {\it intensive} quantity (i.e.~it does not scale with $N$); ii) the physical flux is $\hat{\Phi}=\Phi (\aop + \adop)$. The Peierls phase introduced in Eq.~\eqref{eq: initial hamiltonian} is necessary to satisfy the gauge principle in the presence of magnetic fields and, more in general, in any theory beyond the electrical dipole approximation~\cite{wiese2013lattice,savasta2021gauge, andolina_epjp_2022};  iii) For spinless fermions, TRS operates as a complex conjugation on electronic operators. For photonic operators, TRS changes the sign, i.e. $\aop \rightarrow -\aop$, in order to reverse the direction of the magnetic field. If all hopping terms $t_{j,j+1}$ are real, then the total Hamiltonian given in Eq.~\eqref{eq: initial hamiltonian} is invariant under TRS.  

\section{Mean-field theory and photon condensation criterion}

To the end of studying the possible emergence of photon condensation, we approximate the ground-state of $\hat{\cal H}$ as a product state of the form $\ket{\Psi} = \ket{\psi}\ket{\phi}$, where  $\ket{\psi}$ and $\ket{\phi}$ are the matter and light quantum states \cite{emary_brandes,Chiriaco,Passetti,Eckhard,Amelio_PRB_2021}, respectively.
In the thermodynamic limit, the photonic state can be considered a coherent state $\aop \ket{\phi_{\alpha}} =  \alpha \sqrt{N} \ket{\phi_{\alpha}}$. Photon condensation occurs when the photonic order parameter $\alpha \equiv \langle \phi_{\alpha}| \aop|\phi_{\alpha}\rangle/\sqrt{N}$ acquires a non-zero value in the thermodynamic limit. From now on, we will take $\alpha \in {\mathbb R}$, without loss of generality. We hasten to emphasize that our order parameter $\alpha$ is not plagued by any gauge ambiguity~\cite{Stokes_PRL} since it physically corresponds to the magnetic flux---defined as the expectation value of $\hat{\Phi}$ over the ground state $\ket{\phi_{\alpha}}$, i.e.~$2\alpha \Phi \sqrt{N} = \langle \phi_\alpha|\hat{\Phi}|\phi_\alpha\rangle$---which is measurable e.g. via SQUID or NV-center magnetometry~\cite{casola_natrevmater_2018,barry_RMP_2020}.

We now consider the following mean-field matter Hamiltonian, describing only the electronic degrees of freedom:
\begin{equation}\label{eq:mean-field Hamiltonian per plaquette}
\frac{\hat{\mathcal{H}}_{\rm MF}(\alpha)}{N} \equiv \frac{\mel{\phi_{\alpha}}{\mathcal{\hat{H}}}{\phi_{\alpha}}}{N} = \omega_{\rm c} \alpha^2 +\frac{1}{N} \sum^N_{k=1} \hat{h}_{{\rm e},k} (\alpha)~,
\end{equation}
where
\begin{equation}\label{eq:Hmatk}
\hat{h}_{{\rm e},k} (\alpha)=  { -}  \sum_{j = 0}^{n_{\rm s} -1} \left( t_{j, j+1} e^{-i 2  \lambda \alpha  } \hat{c}_{j,k}^\dagger \hat{c}_{j+1,k} + {\rm H.c.} \right)~.
\end{equation}
Notice that $\hat{h}_{{\rm e},k}(0)$ coincides with the single-molecule Hamiltonian $\hat{h}_{{\rm e},k}$ in the absence of cavity, which has been defined above in Eq.~(\ref{eq:single_molecule_Hamiltonian_no_cavity}). The quantity $\hat{\mathcal{H}}_{\rm MF}(\alpha)/N$ in Eq.~(\ref{eq:mean-field Hamiltonian per plaquette}) can be interpreted as an effective mean-field matter Hamiltonian per molecule.
 
We denote by the symbol $\ket{\varphi_{l}(\alpha)}_k$ and $\epsilon_l (\alpha)$, with $l=0,1,2,\ldots,n_{\rm s}-1$, the eigenstates and the corresponding eigenvalues of the Hamiltonian $\hat{h}_{{\rm e},k} (\alpha)$: $\hat{h}_{{\rm e},k} (\alpha)\ket{\varphi_{l}(\alpha)}_k = \epsilon_l (\alpha)\ket{\varphi_{l}(\alpha)}_k$. The spectrum of $ \hat{h}_{{\rm e},k}(\alpha)$ does not depend on $k$ since all molecules are identical.

A generic many-body eigenstate of the effective Hamiltonian $\hat{\mathcal{H}}_{\rm MF}(\alpha)/N$ can be written as~\cite{antisymmetry} $\ket{\psi_n(\alpha)}=\prod^N_{k=1} \ket{\varphi_{l_{k,n}}(\alpha)}_k$. Here, $l_{k,n}$ is a discrete index that specifies which single-particle state is occupied for the $k$-th molecule and $n = 0, 1,2, \ldots, n^N_{\rm s}-1$ is an integer.
In particular, the many-body ground state (i.e.~$l_{k,0}=0~\forall k$) is given by $\ket{\psi_0(\alpha)}=\prod^N_{k=1} \ket{\varphi_{0}(\alpha)}_k$. Finally, we introduce the energy per molecule $\bar{\epsilon}_n(\alpha)=\sum^N_{k=1} \epsilon_{l_{k,n}} (\alpha)/N$. The ground-state energy is $\bar{\epsilon}_0(\alpha)=\epsilon_{0} (\alpha)$. We emphasize that the quantities $\bar{\epsilon}_n(\alpha)$ and $\epsilon_l (\alpha)$ do not include the electromagnetic energy $\omega_{\rm c}\alpha^2$.

At the onset of the phase transition, $\alpha$ is a small parameter and the mean-field Hamiltonian can be therefore expanded in a power series of $\alpha$, retaining only terms up to ${\cal O}(\alpha^2)$. 
Using the magnetization operator derived in Appendix \ref{subsec: sm magnetization operator}, 
\begin{align}\label{eq: Magnetization operator}
\hat{M}_z(\alpha)= { -}\frac{\lambda }{B_0 \sqrt{N}} \sum_{k, j}\left(  i t_{j,j+1} e^{-i 2 \lambda \alpha} \hat{c}_{j,k}^\dagger \hat{c}_{j+1,k} + {\rm H.c.} \right)~,
\end{align}
we get
\begin{eqnarray}
    \label{eq: Hamiltonian Mp and Md expansion}
    \frac{\hat{\mathcal{H}}_{\rm MF}(\alpha)}{N} = \omega_{\rm c} \alpha^2 + \hat{\mathcal{H}}_0 + \hat{\mathcal{M}}_{\rm p} (0) \alpha + \frac{1}{2 } \hat{\mathcal{M}}_{\rm d} (0) \alpha^2~,
\end{eqnarray}
where $\hat{\mathcal{H} }_0 = \hat{\cal H}_{\rm e}/N$, and we have introduced the paramagnetic $\hat{\mathcal{M}}_{\rm p} (\alpha) =\sum^N_{k=1} \partial_\alpha \hat{h}_{{\rm e},k} (\alpha)/N= - 2 B_0   \hat{M}_z (\alpha)/\sqrt{N}$ and diamagnetic $\hat{\mathcal{M}}_{\rm d} (\alpha) = \partial_\alpha \hat{\mathcal{M}}_{\rm p} (\alpha)$ contributions to the magnetic moment operator. We emphasize that since $ B_0\sim 1/\sqrt{N}$ and $\hat{M}_z (\alpha)\sim N$, both $\hat{\mathcal{M}}_{\rm p} (\alpha)$ and $\hat{\mathcal{M}}_{\rm d} (\alpha)$ are {\it intensive} quantities. In Appendix~\ref{subsec: sm magnetization operator} we show that $\hat{\mathcal{M}}_{\rm p} (\alpha)$ is proportional to the paramagnetic current operator.

As detailed in Appendix~\ref{subsec: condensation criteria}, the conditions for photon condensation to take place can be divided in two classes:

(i) If matter, decoupled from light, displays a non-zero {\it paramagnetic} magnetization,  $\mel{\psi_0(0)}{\hat{\mathcal{M}}_{\rm p} (0)}{\psi_0(0)} \neq 0$---where $\ket{\psi_0(0)}$ is the ground state of $\hat{\mathcal{H}}_0$ with eigenvalue $\bar{\epsilon}_0(0)$---the coupled light-matter system is always in a photon condensate state for any value of the light-matter coupling $\lambda$, and the real ground state energy shift is linear in $\alpha$. The physical reason for this phenomenon is that, since $\mel{\psi_0(0)}{\hat{\mathcal{M}}_{\rm p} (0)}{\psi_0(0)} \neq 0$, the matter ground state $\ket{\psi_0(0)}$ carries a persistent current, which, in turn, creates a non-zero magnetic field in the cavity. This self-generated field corresponding to a finite photonic displacement is the manifestation of photon condensation. Persistent ground-state currents in our planar molecules can be obtained, for example, by applying an external classical magnetic field, which yields a finite $\Theta$  (see Appendix~\ref{subsec: sm external field}),  explicitly breaking TRS. From now on, we will focus only on $\Theta=0$;

(ii) Conversely, as further discussed in Appendix~\ref{subsec: condensation criteria}, if there are no circulating currents in the uncoupled matter system, i.e.~if $\Theta=0$ and $\mel{\psi_0(0)}{\hat{\mathcal{M}}_{\rm p} (0)}{\psi_0(0)} = 0$, photon condensation occurs if and only if
\begin{equation}
\label{eq: photon condensation criteria}
\chi_{\rm M} \geq \omega_{\rm c}~,
\end{equation}
where $\chi_{\rm M} = - \mel{\psi_0(0)}{\hat{\mathcal{M}}_{\rm d} (0)}{\psi_0(0)} / 2 - \chi_{\rm p}(0) / 2$ is the magnetic susceptibility, and
$\chi_{\rm p}(0)= -2 \sum_{n \neq 0} |\langle{\psi_n(0)}|{\hat{\mathcal{M}}_{\rm p} (0)}|{\psi_0(0)}\rangle|^2 /[\bar{\epsilon}_n(0) - \bar{\epsilon}_0(0)]$ is the static paramagnetic susceptibility.   
The above expression for $\chi_{\rm M}$ reveals a link with Van Vleck paramagnetism, as further discussed in Appendix~\ref{subsec: sm Van Vleck}. When two states are degenerate, i.e., $\bar{\epsilon}_1(0) \to \bar{\epsilon}_0(0)$, the magnetic susceptibility tends to $+\infty$, automatically satisfying the photon condensation criterion.

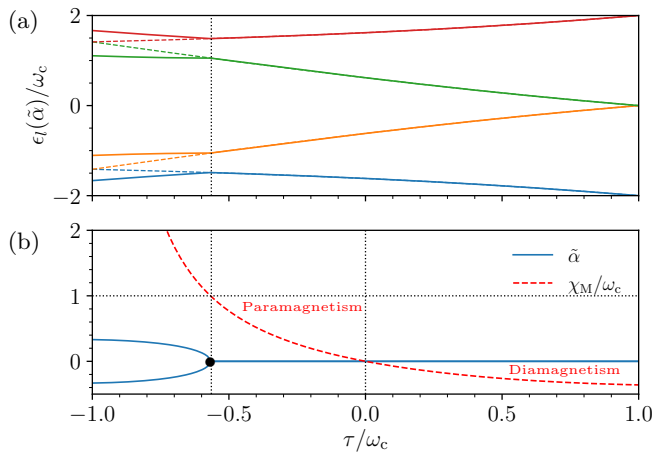
\begin{figure}[t]
    \centering
\begin{overpic}[width=\columnwidth]{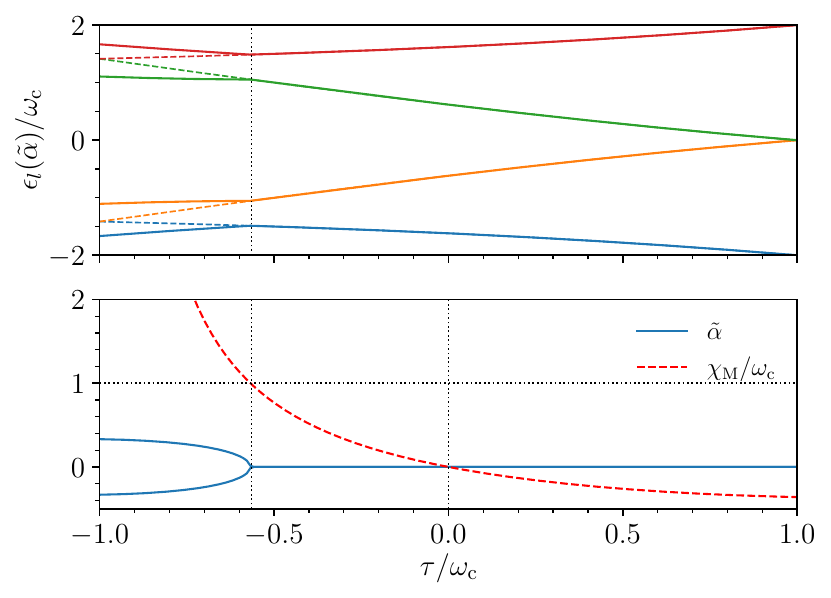}\put(-1,67){(a)}\put(-1,33){(b)}\put(76,13.7){\rm \tiny {\color{red}Diamagnetism}}\put(35,23.5){\rm \tiny {\color{red}Paramagnetism}}\put(29.2,14.7){$\bullet$}\end{overpic}
\caption{(Color online) Photon condensation of molecules in a chiral cavity. Results obtained by setting $\Theta=0$, $\lambda=0.3$, and { $t=\omega_{\rm c}$}. (a) Eigenvalues $\epsilon_l(\tilde{\alpha})$ of the single-molecule Hamiltonian $\hat{h}_{{\rm e},k}$ plotted as functions of $\tau/\omega_{\rm c}$. Solid lines denote results for $\lambda=0.3$. Dashed lines denote results in the absence of the cavity. The vertical dotted line marks the critical value of $\tau$ at which a transition to a photon condensate state occurs. (b) The photon condensate order parameter $\tilde{\alpha}$ is plotted as a function of $\tau$. When $\tau$ reaches a critical value, $\tilde{\alpha}$ becomes finite, signaling a quantum phase transition to a photon condensate state. \label{fig: condensation criteria} 
}
\end{figure}

Eq.~(\ref{eq: photon condensation criteria}) is the most important result of this Letter. Despite it was derived for a toy-model molecular Hamiltonian, we believe that its range of validity is much more ample. For example, the inclusion of electron-electron interactions is not expected to modify Eq.~(\ref{eq: photon condensation criteria}) but to dramatically alter  the dependence of $\chi_{\rm M}$ on the microscopic molecular parameters. Also, the actual details of the cavity will certainly matter but deviations from our toy-model temporally-chiral cavity hosting a spatially-uniform fluctuating ${\bm B}$ field can be easily encoded in the right-hand side of the inequality (\ref{eq: photon condensation criteria}), changing $\omega_{\rm c}$ into a more complicated electromagnetic form factor. Our criterion (\ref{eq: photon condensation criteria}) provides guidance in the experimental search for photon condensation in polaritonic chemistry~\cite{fregoni_acsnano_2022}, emphasizing that the quest for this exotic state of matter should focus on the combination between molecular systems with a positive orbital magnetic susceptibility $\chi_{\rm M}$ and chiral cavities.

\section{Variational theory of the photon condensate state}

The perturbative approach for $\alpha \ll 1$ described so far served only to reach the criterion (\ref{eq: photon condensation criteria}) for  photon condensation. If one is interested in the actual calculation of the order parameter $\alpha$ and the spectra ${\bar \epsilon}_l$ of the coupled light-molecular system as functions of the microscopic parameters of the model, a non-perturbative approach is needed.  To this end, we fix the parameter $\alpha$ by imposing that the optimal value $\tilde{\alpha}$ yields a minimum of the mean-field ground-state energy functional ${\cal E}(\alpha)\equiv\bra{\psi_0 (\alpha)} \hat{\cal H}_{\rm MF}(\alpha) \ket{\psi_0 (\alpha) }/N$. Hence, we determine $\tilde{\alpha}$ by imposing that $\partial_\alpha {\cal E}(\alpha)|_{\alpha=\tilde{\alpha}}=0$:
\begin{equation}\label{eq: solutions for alpha}
\left.\partial_\alpha {\cal E}(\alpha)\right|_{\alpha=\tilde{\alpha}}=2\omega_{\rm c} \tilde{\alpha}+\mel{\psi_0(\tilde{\alpha})}{\hat{\mathcal{M}}_{\rm p} (\tilde{\alpha})}{\psi_0(\tilde{\alpha})}=0~.
\end{equation}
The solution of Eq.~\eqref{eq: solutions for alpha} not only determines whether the system is in a normal ($\tilde{\alpha}=0$) or photon condensate ($\tilde{\alpha}\neq 0)$ phase but also yields $\tilde{\alpha}$ as a function of the microscopic parameters $\tau$, $\Theta$, and $\lambda$.  If $\Theta=0$, the photon condensate state ($\tilde{\alpha}\neq 0)$ spontaneously breaks TRS due the appearance of a finite magnetic field and circulating currents. Conversely, in the case $\Theta\neq0$, the system is not invariant under TRS to begin with. Results for $\tilde{\alpha}$ as a function of $\tau/\omega_{\rm c}$ and $\Theta$ for $\lambda=0.3$ have been reported in Fig.~\ref{fig: 3d_fig}.

Figure~\ref{fig: condensation criteria}(a) shows the molecular spectra $\epsilon_l(\tilde{\alpha})$ as functions of $\tau / \omega_{\rm c}$. The dashed lines describe the molecular spectra in the absence of cavity (i.e.~for $\lambda = 0$), while the solid lines describe the case of a finite light-matter coupling ($\lambda = 0.3$). The vertical dotted line marks the critical value of $\tau$ beyond which a transition to a photon condensate state occurs. We clearly see that at this value of $\tau$ the spectra are largely affected by the cavity. Fig.~\ref{fig: condensation criteria}(b) shows the order parameter $\tilde{\alpha}$ (solid blue line) and the orbital magnetic susceptibility $\chi_{\rm M}$ (red dashed line) as functions of $\tau / \omega_{\rm c}$. In agreement with Eq.~\eqref{eq: photon condensation criteria}, photon condensation (i.e.~$\tilde{\alpha}\neq 0$) occurs when $\chi_{\rm M}/\omega_{\rm c} > 1$. Diamagnetism (paramagnetism) corresponds to $\chi_{\rm M}<0$ ($\chi_{\rm M}>0$). Note that $\chi_{\rm p}(0) \to -\infty$ (therefore yielding $\chi_{\rm M} \to +\infty$) due to the degeneracy $\epsilon_1(0)=\epsilon_0(0)$ shown in Fig.~\ref{fig: condensation criteria}(a) (see blue and orange dashed lines at {$\tau=-\omega_{\rm c}$}).

\section{Polaritons}

Measuring directly the molecular spectrum $\epsilon_l(\tilde{\alpha})$ in the presence of the cavity is of course possible but challenging. Light-matter interactions yield also polaritons, i.e.~hybrid light-matter collective modes, which can be measured in a variety of ways, including scanning probe methods~\cite{Basov_Science_2016,Low_NatureMater_2017,Basov_Nanophotonics_2021,Plantey2021}. In order to find polaritons in our system, we need to study Gaussian fluctuations around the mean-field state described by the Hamiltonian (\ref{eq:mean-field Hamiltonian per plaquette}). To this aim, we write the photon operators as $\aop \rightarrow \tilde{\alpha} \sqrt{N}+\delta \aop $, where $\delta \aop$ describes a zero-average fluctuation around the mean-field solution $\tilde{\alpha}\sqrt{N}$, which has been described above. We then introduce the collective ``bright mode'' creation operator~\cite{hopfield_physrev_1958,savona_prb_1994,ciuti_prb_2005,gerace_prb_2007,carusotto_pra_2008,hagenmuller_prb_2010}, 
\begin{equation}\label{eq:bmode}
\hat{b}_l^\dagger \equiv \frac{1}{\sqrt{N}}  \sum^N_{k=1} \hat{\varphi}^\dagger_{l,k} \hat{\varphi}_{0,k}
\end{equation}
with $l>0$, where $\hat{\varphi}^\dagger_{l,k} $ ($\hat{\varphi}_{0,k}$) creates (destroys) an electron in the  eigenstate $\ket{\varphi_{l}(\tilde{\alpha})}_k$ ($\ket{\varphi_{0}(\tilde{\alpha})}_k$). 
The bright mode collective operator creates an electron-hole transition with a finite electrical dipole moment by annihilating an electron in the ground state of each molecule and creating an electron in an excited state $l>0$ with energy $\epsilon_l(\tilde{\alpha})$. The associate transition energy is $\epsilon_l(\tilde{\alpha})-\epsilon_0(\tilde{\alpha})$. In the thermodynamic $N\to \infty$, $\hat{b}^\dagger_l$ behaves as a quasi-bosonic operator~\cite{hopfield_physrev_1958,savona_prb_1994,ciuti_prb_2005,gerace_prb_2007,carusotto_pra_2008,hagenmuller_prb_2010}, i.e.~$[\hat{b}_m, \hat{b}^\dagger_l]\approx \delta_{m,l}$. Further details are given in Appendix~\ref{subsec: sm Bosonization and polaritons study}.

Hence, in the thermodynamic limit, and expanding the shifted Hamiltonian up to the second order in the photonic fluctuations $\delta\aop$ and in the bright mode bosonic operators $ b_l^\dagger$, we can write an approximate {\it polaritonic} Hamiltonian $\hat{\cal H}_{\rm pol}$, describing the lowest excited states of the coupled cavity-molecular system:
\begin{eqnarray}
\label{eq:Hpolaritons}
\hat{\cal H}_{\rm pol} &=& \omega_{\rm c} \delta \adop \delta \aop 
+ \sum_{l=1}^{n_{\rm s} - 1} [\epsilon_{l}(\tilde{\alpha}) -\epsilon_{0}(\tilde{\alpha})]\hat{b}_l^\dagger \hat{b}_l \nonumber\\
&+& \frac{1}{2} (\delta \aop + \delta \adop) \sum_{l=1}^{n_{\rm s} - 1} \left[ \mathcal{M}_{\rm p}^{l,0} ( \tilde{\alpha} ) \hat{b}_l^\dagger + \mathcal{M}_{\rm p}^{0,l} (\tilde{\alpha}) \hat{b}_l \right] \nonumber \\
&+& \frac{1}{8} \mathcal{M}_{\rm d}^{0, 0} ( \tilde{\alpha} ) (\delta \aop + \delta \adop)^2~,
\end{eqnarray}
where $\mathcal{M}_{\gamma}^{l, m} (\tilde{\alpha} )=\sum^N_{k=1}  \tensor[_{k}]{\langle}{}  {\varphi_{l}(\tilde{\alpha} )}| 
\hat{\cal M}_{\gamma}(\tilde{\alpha} )
\ket{\varphi_{m}(\tilde{\alpha} )}_k $ with $\gamma={\rm p} ,{\rm d}$. Note that $\mathcal{M}_{\rm p}^{0, 0} (\tilde{\alpha} )=\langle \psi_0(\tilde{\alpha})|\hat{\cal M}_{\rm p}(\tilde{\alpha})|\psi_0(\tilde{\alpha})\rangle$ and $\mathcal{M}_{\rm d}^{0, 0} (\tilde{\alpha} )=\langle \psi_0(\tilde{\alpha})|\hat{\cal M}_{\rm d}(\tilde{\alpha})|\psi_0(\tilde{\alpha})\rangle$. Finally, the polariton frequencies can be derived by diagonalizing the Hopfield matrix $\Xi$, i.e.~the matrix that represents the quadratic polaritonic Hamiltonian in Eq.~(\ref{eq:Hpolaritons}). Further details are reported in Appendix~\ref{subsec: sm Bosonization and polaritons study}.

\begin{figure}[t]
\centering
\includegraphics[width=\linewidth]{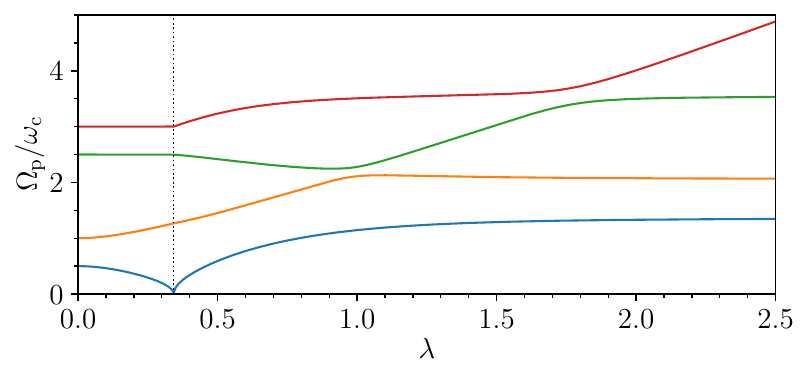}
\caption{(Color online) Polariton softening at finite $\lambda$. Results in this figure have been obtained by setting with $\Theta=0$, $\tau = -0.5 \omega_{\rm c}$, and $t=\omega_{\rm c}$. The four polariton energies $\Omega_{\rm p}$ are plotted as functions of $\lambda$. The most important feature of this panel is the softening of the lowest polariton mode, which occurs at the quantum phase transition to a photon condensate state. \label{fig:polaritons_1}}
\end{figure}
%


Figure~\ref{fig:polaritons_1} shows the four polariton frequencies $\Omega_{\rm p}$ as functions of $\lambda$ and for a fixed value of $\tau$. We clearly see that, at $\lambda=\lambda_{\rm c}$, the lowest polariton mode softens, signaling that the transition to a photon condensate state is a second-order quantum phase transition. When $\lambda$ exceeds a critical value $\lambda_{\rm c}$, $\tilde{\alpha}$ increases from zero to a finite value. Physically, this means that at $\lambda>\lambda_{\rm c}$, a magnetic flux appears spontaneously.

The physics discussed so far does not require large values of the light-matter coupling $\lambda$. Although reaching the ultra-strong coupling regime~\cite{kockum_naturereviewsphysics_2019,fregoni_acsnano_2022,VacuumReview,Genet_PT_2021} is possible in a variety of condensed matter and quantum chemistry setups, it is highly desirable to have toy models and realistic systems where photon condensation occurs in the weak-coupling $\lambda\to 0$ limit. In Appendix~\ref{subsec: sm Bosonization and polaritons study}, we show that our model displays such a pleasant feature, provided that one chooses  $\tau=-\omega_{\rm c}$. Indeed, for photon condensation to occur is more important to hunt for molecular systems with a large value of $\chi_{\rm M}>0$ and design cavities with a suitable electromagnetic vacuum structure so that $\chi_{\rm M}/\omega_{\rm c}>1$ rather than achieving ultra-strong coupling.


In summary, we have shown that photon condensation can occur also in molecular systems (and not only in extended electronic systems~\cite{nataf_prl_2019,andolina_prb_2020,guerci_prl_2020}) provided that magnetic effects beyond the electrical dipole approximation are taken into account. The recipe for achieving it is encoded in the simple and elegant criterion we derived in Eq.~(\ref{eq: photon condensation criteria}). One needs to find molecules with a large and positive orbital magnetic susceptibility $\chi_{\rm M}$ and place them in cavities hosting a significant magnetic component of the electromagnetic field. In order to grasp the essential physics, we have deliberately analyzed, for the sake of simplicity, single-electron toy-model molecules placed inside a temporally-chiral cavity with a uniform magnetic field. We hope that our results will stimulate future work on real molecules loaded into more complex chiral cavities~\cite{hubener_naturematerials_2021}, which can be studied with recently developed {\it ab initio} numerical approaches~\cite{Haugland_PRX_2020}.
 Finally, ultra-strong magnetic coupling between magnons and a planar superconducting resonator \cite{ghirri_apl_2015} has been recently demonstrated \cite{ghirri_arxiv_2302.00804}. This could be a promising platform to test our findings. 
Results similar to ours, which show the importance of a cavity with a significant magnetic component, have been also found in a two-leg ladder model: see Ref.~\cite{Bacciconi}. 

\section{Acknowledgements}

This work was supported by: i) the European Union's Horizon 2020 research and innovation programme under grant agreement no.~881603 - GrapheneCore3; ii) the University of Pisa under the ``PRA - Progetti di Ricerca di Ateneo" (Institutional Research Grants) -  Project No.~PRA\_2020-2021\_92 ``Quantum Computing, Technologies and Applications''; and iii) the Italian Minister of University and Research (MUR) under the ``Research projects of relevant national interest  - PRIN 2020''  - Project no.~2020JLZ52N, title ``Light-matter interactions and the collective behavior of quantum 2D materials (q-LIMA)''. This work has been partially funded by European Union (NextGeneration EU), through the MUR-PNRR project SAMOTHRACE (ECS00000022). F.M.D.P. was supported by the PNRR MUR project PE0000023-NQSTI. P.J.H. acknowledges support by the Gordon and Betty Moore Foundation’s EPiQS Initiative through grant GBMF9463, the Fundacion Ramon Areces, and the ICFO Distinguished Visiting Professor program. F.H.L.K. acknowledges financial support from  the ERC TOPONANOP  (726001), the Government of Catalonia through the SGR grant, the Spanish Ministry of Economy and Competitiveness through the Severo Ochoa Programme for Centres of Excellence in R\&D (Ref. SEV-2015-0522) and Explora Ciencia (Ref. FIS2017- 91599-EXP), Fundacio Cellex Barcelona, Generalitat de Catalunya through the CERCA program, the Mineco grant Plan Nacional (Ref. FIS2016-81044-P), and the Agency for Management of University and Research Grants (AGAUR) (Ref. 2017-SGR-1656). This research has received the financial support of the “Ministerio de Ciencia e Innovación” (Real Decreto 768/2022, de 20 de Septiembre, Quantica Valley project). S.S. acknowledges support by the Army Research Office (ARO) through grant No. W911NF1910065. It is a great pleasure to thank Z. Bacciconi, T. Chanda, G. Chiriac\`o, M. Dalmonte, G. Mazza, M. Schir\'o and  E. Ronca for useful discussions.

\appendix

\section{\label{subsec: molecules}Derivation of the effective electronic Hamiltonian}

A (single) molecule is composed by mutually interacting electrons and nuclei. Their Hamiltonian $\hat{H}$ includes their respective kinetic energies and all Coulomb interactions among them (electron-electron, electron-nucleus, and nucleus-nucleus). In the non-relativistic limit, this Hamiltonian reads:

\begin{equation}
\hat{H} = \hat{T}_{\rm N} + \hat{H}_{\rm e}(\bm{R}) + \hat{V}_{\rm NN},
\end{equation}

where:

\begin{align*}
\hat{T}_{\rm N} &= -\sum_{I=1}^{Z}\frac{\hat{\bm{P}}^2_I}{2M_I}  ~,\\
\hat{V}_{\rm NN} &= \frac{1}{2}\sum_{I\neq J}^{Z} \frac{Z_I Z_Je^2}{|\bm{R}_I - \bm{R}_J|}~,\\
\hat{H}_{\rm e}(\bm{R}) &= - \sum_{i=1}^{N} \frac{\hat{\bm{p}}^2_i} {2m}+ \frac{1}{2}\sum_{i\neq j}^{N} \frac{e^2}{|\bm{r}_i - \bm{r}_j|}+\\
&-\sum_{i=1}^{N}\sum_{I=1}^{Z} \frac{Z_Ie^2}{|\bm{r}_i - \bm{R}_I|}~.
\end{align*}

In these equations, $\bm{r}_i$ ($\hat{\bm{p}}_i$) and $\bm{R}_I$ ($\hat{\bm{P}}_I$) are the position (momentum) of the $i$-th electron and the $I$-th ion, respectively.

Given the large difference in mass between electrons and nuclei, $M_I\gg m$, the dynamics of electrons in a molecule is substantially faster than that of the nuclei.
Due to this separation of time scale, we can invoke the Born-Oppenheimer approximation, which assumes that the nuclear motion and the electronic motion can be decoupled and, therefore, the total wavefunction of the system $\Psi(\bm{r},\bm{R})$ is assumed to be a product of an electronic wavefunction $\psi(\bm{r};\bm{R})$, which depends on the nuclear positions, and a nuclear wavefunction $\chi(\bm{R})$,  $\Psi(\bm{r},\bm{R}) = \chi(\bm{R})\psi(\bm{r};\bm{R})$.

Within this approximation,  when we deal with electrons we can ignore the motion of the nuclei, and assume that they see a fixed arrangement of charge. Hence, the electronic wavefunction satisfies the resulting electronic Schr\"odinger equation:

\begin{equation}
\hat{H}_{\rm e}(\bm{R}) \psi(\bm{r};\bm{R}) = E_{\rm e}(\bm{R}) \psi(\bm{r};\bm{R}),
\end{equation}

where $E_{\rm e}(\bm{R})$ is the electronic energy corresponding to a given nuclear configuration $\bm{R}$. The function $\psi(\bm{r};\bm{R})$ can be thought of as the ground state solution of the electronic Schr\"odinger equation for fixed nuclear positions. In this expression, the electronic Hamiltonian $\hat{H}_{\rm e}(\bm{R})$ is only parametrically dependent on the static nuclear configuration defined by $\bm{R}$, which can be treated as classical parameters.

Lastly, we project the electronic Hamiltonian $\hat{H}_{\rm e}(\bm{R})$ on a basis of localized wavefunctions in order to get a simplified tight binding Hamiltonian. Given $\hat{H}_e(\bm{R})$, we can project it onto the localized basis ${|j\rangle}$ by computing the matrix elements of $\hat{H}_e(\bm{R})$ in this basis:

\begin{equation}
H_{j,j^\prime} = \langle j|\hat{H}_e(\bm{R})|j^\prime \rangle.
\end{equation}

Here, $H_{j,j^\prime}$ represents the matrix element of the electronic Hamiltonian $\hat{H}_e(\bm{R})$ projected on localized states $|n\rangle$ and $|m\rangle$. If the states $|j\rangle$ and $|j^\prime \rangle$ are well-localized, $H_{j,j^\prime}$ would be zero for most pairs $(j,j^\prime)$, leading to a sparse Hamiltonian matrix - a key feature of tight-binding models.

In general, the elements $H_{j,j^\prime}$ will generally include terms related to the on-site energy of a particle in state $|j\rangle$, and terms related to the hopping of a particle from state $|j\rangle$ to state $|j^\prime\rangle$. Hence, we obtain a tight-binding Hamiltonian in real space for electrons:

\begin{equation}
\label{eq:HTB}
\hat{H}\approx \sum_{j} \epsilon_j |j\rangle\langle j| -\sum_{j \neq j^\prime} t_{j,j^\prime} |j\rangle\langle j| + \text{H.c.}
\end{equation}

Here $\epsilon_j$ is the on-site energy of an electron on site $j$, $t_{j,j^\prime}$ is the hopping amplitude for an electron to move from the site $j^\prime$ to site $j$. Both $\epsilon_j$ and $t_{j,j^\prime}$ can be determined by computing the appropriate matrix elements of the original Hamiltonian, as mentioned above. In what follows, we assume $\epsilon_j=0$.

In second quantization, we can rewrite Eq.~\eqref{eq:HTB} as:

\begin{equation}
\label{eq:HTB1}
\hat{H}\approx - \sum_{j \neq j^\prime} t_{j,j^\prime} \hat{c}^\dagger_{j} \hat{c}_{j^\prime} + \text{H.c.}
\end{equation}

where $\hat{c}_{j}^\dagger$ and $\hat{c}_{j}$ are the creation and annihilation operators for an electron at site $j$, respectively.
Eq. (1) of the main text is a particular instance of Eq.~\eqref{eq:HTB1}, providing the addition of a further index $k$ to denote different molecules. 

\section{\label{subsec: sm magnetization operator}Derivation of the magnetization and current operators}

In this Section, we derive the explicit form of the magnetization operator for electrons hopping in a single ``molecule'' in a second-quantized fashion.
The position operator $\hat{\bm r}_k$, associated with the $k$-th  molecule, can be written in terms of electron creation and annihilation operators, $\hat{c}^\dagger_{j,k}$ and $\hat{c}_{j,k}$ for an electron roaming on a polygon with $n_{\rm s}$ sides:
\begin{equation}
\hat{\bm r}_k=\sum^{n_{\rm s}-1}_{j=0} {\bm r}_j\hat{c}^\dagger_{j,k} \hat{c}_{j,k}~.
\end{equation}
Here, ${\bm r}_j= d_r [\cos(\gamma_j),\sin(\gamma_j)]^{\rm T}$ is the position of the $j$-th site measured from the center of the molecule, $d_r$ is the distance of the site from the center and $\gamma_j=2\pi j/n_{\rm s}$ is the angle subtended between the position vector of the first site and the one of the $j$-th site.
The magnetic moment of a single molecule is defined in terms of the position operator as \cite{Jackson}
\begin{equation}
\label{eq:Mk}
\hat{\bm M}_k
=\frac{1}{2}   \hat{\bm r}_k \times \Big(-\frac{e}{c} \dot{\hat{\bm r}}_k\Big)~.
\end{equation}
We use the Heisenberg equation of motion to find an expression for the velocity operator $\dot{\hat{\bm r}}_k$,
\begin{equation}
\dot{\hat{\bm r}}_k= i [\hat{\cal H},\hat{\bm r}_k]~.
\end{equation}
Replacing this result in the definition of the magnetic moment (Eq.~\eqref{eq:Mk}) we get
\begin{equation}
\label{eq:Mk1}
 \hat{\bm M}_k=\frac{-e}{c}\frac{i }{2}    \hat{\bm r}_k \times  [\hat{\cal H},\hat{\bm r}_k]~. 
\end{equation}
We now consider the magnetic moment along the ${\bm e}_z$ direction,  $\hat{M}_{z,k}$, and we use the explicit form of the total Hamiltonian $\hat{\cal H}$ in Eq.~(\ref{eq:Mk1}). We find
\begin{align}\label{eq:Sm_BMz_k-th_molecule}
    \hat{M}_{z,k} &= -\frac{\lambda }{B_0 \sqrt{N}} \times \nonumber \\ \times \sum_{j = 0}^{n_{\rm s} - 1}
   & \left[  i t_{j,j+1} e^{-i \frac{\lambda}{\sqrt{N}} (\hat{a}+\hat{a}^\dagger)} \hat{c}_{j,k}^\dagger \hat{c}_{j+1,k} + {\rm H.c.} \right]~.
\end{align}
In the presence of $N$ identical molecules, the total magnetization operator is expressed as 
\begin{equation}
\hat{M}_z=\sum^N_{k=1}\hat{M}_{z,k}~.
\end{equation}
In the mean-field approach described in the main text,  we can trace out the photonic degrees of freedom by projecting the total magnetization onto a coherent state $\ket{\phi_{\alpha}}$, i.e.~onto a state such that $\hat{a}\ket{\phi_{\alpha}}=\alpha\sqrt{N}\ket{\phi_{\alpha}}$.
In the thermodynamic limit ($N \to \infty $), this mean-field procedure reduces to the following formal replacement $\hat{a}\to \alpha\sqrt{N}$. In this limit,  
the total magnetization operator becomes
\begin{equation}\label{eq:SM_B0Mz0}
    \hat{M}_z(\alpha)= \sum^N_{k=1}\hat{M}_{z,k}(\alpha)~,
\end{equation}
where
\begin{equation}\label{eq:SM_B0Mz1}
    \hat{M}_{z,k}(\alpha)=-\frac{\lambda}{B_0  \sqrt{N} } \sum^{n_{\rm s}-1}_{j=0 }
    \left(  i t_{j,j+1} e^{-i2 \lambda \alpha} \hat{c}^\dagger_{j,k} \hat{c}_{j+1,k} + {\rm H.c.} \right)~.
\end{equation}
To map this operator into a more familiar object with the physical dimensions of a magnetic moment, it is necessary to introduce an effective mass $m_{\rm eff}$ for our electrons hopping through the lattice sites: $m_{\rm eff} = 1/(2 |t| A_{\rm p})$, where $A_{\rm p}$ is the area of a plaquette and $t$ is the hopping parameter that we have introduced in the main text. For instance, for a square plaquette of side $d$ ($n_{\rm s}=4$), the effective mass is $m_{\rm eff}=1/(2|t| d^2)$. Introducing the effective mass into the definition of the magnetization operator we find

\begin{align}\label{eq:SM_B0Mz2}
\hat{M}_z(\alpha)=-\frac{e}{2 m_{\rm eff} c}\sum_{k, j}\left( - i \frac{t_{j,j+1}}{n_s |t|} e^{-i 2 \lambda \alpha} \hat{c}_{j,k}^\dagger \hat{c}_{j+1,k} + {\rm H.c.} \right)~,
\end{align}

where the prefactor has the natural form of a magnetic moment $e/(2m_{\rm eff} c)$. 

The total current operator $\hat{J}$ can be derived by starting from a discretized form of the continuity equation for the density operator. The local density $\hat{n}_{j,k}\equiv\hat{c}^\dagger_{j,k} \hat{c}_{j,k}$ obeys the Heisenberg equation of motion
\begin{equation}\label{eq:Heis}
\dot{\hat{n}}_{j,k}= i [\hat{\cal H},{\hat{n}}_{j,k}]~.
\end{equation}
By this comparing this equation to the following discretized continuity equation,
\begin{equation}\label{eq:Cont1}
\dot{\hat{n}}_{j,k}= -\frac{1}{d}\big(\hat{J}_{j,k}-\hat{J}_{j-1,k}\big)~,
\end{equation}
where $\hat{J}_{j,k}$ is the current flowing from site $j$ to site $j+1$ and $d$ is the lattice spacing, we obtain the following expression for the local current,
\begin{equation}\label{eq:Cont2}
{\hat{J}}_{j,k}= d\big(-it_{j,j+1} e^{-i \lambda (\hat{a}+\hat{a}^\dagger)/\sqrt{N}}\hat{c}^\dagger_{j+1,k} \hat{c}_{j,k}{\color{blue} +}{\rm H.c}\big)~.
\end{equation}
The total current $\hat{J}$ is the sum of all local terms, i.e.~$\hat{J}=\sum_{k,j} {\hat{J}}_{j,k}$, and is given by
\begin{equation}\label{eq:Curr1}
{\hat{J}}= d\sum_{k,j}\big({-}it_{j,j+1}e^{-i \lambda (\hat{a}+\hat{a}^\dagger)/\sqrt{N}}\hat{c}^\dagger_{j+1,k} \hat{c}_{j,k}+{\rm H.c}\big)~.
\end{equation}
As usual, in the mean-field approximation, we replace $\hat{a}\to \alpha\sqrt{N}$ obtaining a current operator acting only on the matter degrees of freedom:
\begin{equation}\label{eq:Curr2}
{\hat{J}}(\alpha)= d \sum_{k,j}\big(- it_{j,j+1}e^{-i2 \lambda\alpha }\hat{c}^\dagger_{j+1,k} \hat{c}_{j,k}+{\rm H.c}\big)~.
\end{equation}
As expected, the current operator  ${\hat{J}}(\alpha)$ and the magnetization operator $\hat{M}_z(\alpha)$ in Eq.~\eqref{eq:SM_B0Mz2} are directly proportional to each other. Specifically, $\hat{M}_z(\alpha)=-({e}/{c})  {\hat{J}}(\alpha) [{A_{\rm p}}/({n_s d})]$.

\section{\label{subsec: sm external field}The effect of an external classic field}

As we have seen in the main text, the orbital magnetic response changes sign when $\tau<0$, i.e.~when one of the hopping parameters in the single-molecule Hamiltonian (1) has a different sign with respect to the others (we remind the reader that {$t>0$}). We now explain how one can achieve this frustrated condition on the hoppings by considering the action of an external classical magnetic field. 
Starting from the mean-field Hamiltonian [cfr.~Eq.~(5) in the main text],
\begin{align}
\frac{\hat{\mathcal{H}}_{\rm MF}(\alpha)}{N} &=\frac{\mel{\phi_{\alpha}}{\mathcal{\hat{H}}}{\phi_{\alpha}}}{N} = \omega_{\rm c} \alpha^2 +\frac{1}{N} \sum^N_{k=1} \hat{h}_{{\rm e},k}(\alpha)~,\nonumber\\
\hat{h}_{{\rm e},k}(\alpha)&=- 
\sum_{j = 0}^{n_{\rm s} -1} \left( e^{-i 2  \lambda \alpha  } t_{j, j+1} \hat{c}^\dagger_{j,k} \hat{c}_{j+1,k}+ {\rm H.c.} \right)~, \nonumber
\end{align}
and considering the special case in which all the hopping parameters are equal to $t >0$ but for $t_{0,1} = e^{-i \Theta} t$, we see that the matter Hamiltonian reduces to
\begin{eqnarray}\label{eq:S-Bcl}
\hat{h}_{{\rm e},k} (\alpha)&=&  - \sum_{j = 1}^{n_{\rm s} -1} \left(  t e^{-i 2  \lambda \alpha  }  \hat{c}^\dagger_{j,k} \hat{c}_{j+1,k} + {\rm H.c.} \right) \nonumber \\
&-& \left(  te^{-i \Theta}  e^{-i 2  \lambda \alpha  }  \hat{c}^\dagger_{0,k} \hat{c}_{1,k} + {\rm H.c.} \right)~.
\end{eqnarray}
By applying the unitary transformation $\hat{c}_{0,k}\to e^{-i \Theta}  \hat{c}_{0,k}$ and $ \hat{c}_{j,k}\to e^{-i \Theta (j / n_{\rm s})}  \hat{c}_{j,k}$ for $j\neq 0$,  we get 
\begin{align}
\label{eq:H_ek}
\hat{h}_{{\rm e},k}(\alpha) &= - \sum_{j = 0}^{n_{\rm s} -1} \left(  t e^{-i \Theta/ n_{\rm s}}  e^{-i 2  \lambda \alpha  }  \hat{c}^\dagger_{j,k} \hat{c}_{j+1,k}+ {\rm H.c.} \right)~.
\end{align}
This suggests that a classical magnetic field $\bm{B}_{\rm cl}=B_{\rm cl} \bm{e}_z$ with $B_{\rm cl} = - c \Theta / (e A_{\rm p})$ can be used to change the sign of one of the hopping parameters, thereby paving the way for orbital paramagnetism in our toy model and the occurrence of photon condensation.

The phase $\Theta$ in \eqref{eq:S-Bcl} can be straightforwardly obtained 
by applying the Peierls substitution which describes the orbital effect of $\bm{B}_{\rm cl}$. Using the vector potential in Landau gauge ${\bm A} (\bm{r})=(0,B_{\rm cl}x,0)^{\rm T}$ which generates the static magnetic field $\bm{B}_{\rm cl}$, in a square plaquette with the four sites located at  $(0,0)^{\rm T}$, $(a,0)^{\rm T}$, $(0,a)^{\rm T}$, and $(a,a)^{\rm T}$, one finds that the phase of the link between the sites $j$ and $j=+1$, $\theta_{j, j + 1} = (-e/c) \int_{{\bm r}_j}^{{\bm r}_{j + 1}} {\bm A}({\bm r}) \cdot d^2 {\bm r}$, is not zero only for the link that connects $(a,0)^{\rm T}$ and $(a,a)^{\rm T}$.

{ The Hamiltonian in Eq.~\eqref{eq:H_ek} can be diagonalized by performing a discrete Fourier transformation (DFT) on the index $j$ of the creation and annihilation operators. Let us denote the transformed operators as $\hat{c}_{q,k}$, which is a function of momentum $q$ and molecule index $k$. The DFT will be as follows:

\begin{equation}
\hat{c}_{j,k} = \frac{1}{\sqrt{n_{\rm s}}} \sum_{q} e^{i 2 \pi j q / n_{\rm s}} \hat{c}_{q,k}~,
\end{equation}

and its Hermitian conjugate

\begin{equation}
\hat{c}^\dagger_{j,k} = \frac{1}{\sqrt{n_{\rm s}}} \sum_{q} e^{-i 2 \pi j q / n_{\rm s}} \hat{c}^\dagger_{q,k}~.
\end{equation}
 
For a system with $n_{\rm s}$ sites, the allowed momentum states are given by $q = 0, 1, 2, ..., n_{\rm s} - 1$.
Substituting the DFT of $\hat{c}_{j,k}$ and $\hat{c}^\dagger_{j,k}$ into the Hamiltonian $\hat{h}_{{\rm e},k}(\alpha)$, and
performing the sum over $j$ leads to

\begin{align}
\hat{h}_{{\rm e},k}(\alpha)&= - 2t \sum_{q} \cos \left(\Theta/ n_{\rm s} + 2 \lambda \alpha + 2 \pi q / n_{\rm s} \right) \hat{c}^\dagger_{q,k} \hat{c}_{q,k}~.
\end{align}

The ground state of this Hamiltonian, when it contains a single electron, is denoted by $\hat{c}^\dagger_{\tilde{q},k}\ket{0}$, where $\tilde{q}$ denotes the momentum corresponding to the lowest energy. In this notation, $\ket{0}$ represents the vacuum state, the state in which there are no electrons.

The energy of this ground state is given by
{ $\epsilon_{0}(\alpha)=\min_{q}[-2t \cos \left(\Theta/ n_{\rm s} + 2 \lambda \alpha + 2 \pi q / n_{\rm s} \right) ]$}.
%
The total energy of the system, $\cal{E}(\alpha)$, is calculated by summing the electronic energy and the energy of the cavity, expressed as ${\cal{E}}(\alpha)={\epsilon_{0}(\alpha)}+\omega_{\rm c} \alpha^2$. The minimum total energy is found by setting the derivative of the total energy with respect to $\alpha$ to zero, at $\alpha=\tilde{\alpha}$, leading to the following condition:

\begin{align}
2\omega_{\rm c}\tilde{\alpha}+\frac{d \epsilon_{0}(\alpha)}{d \alpha}\Big|_{\alpha=\tilde{\alpha}}=0~.
\end{align}

An important observation is that, when $\Theta=0$,  $\tilde{\alpha}=0$ is the only solution. This means that if all the hopping parameters are taken to be equal, the phenomenon of photon condensation does not occur in this model.

}

\section{\label{subsec: condensation criteria}Condensation criteria}
In this Section, we derive a general criterion to study the onset of photon condensation. Here, we will consider the energy functional $E[\alpha,\psi]\equiv\langle\psi| \hat{\mathcal{H}}_{\rm MF}(\alpha)|\psi\rangle/N$ as a function of the photonic order parameter $\alpha$ and an arbitrary trial wave-function $|\psi\rangle$. 
By analyzing the function ${\cal E}(\alpha)={\rm min}_{\psi}E[\alpha,\psi]=\langle\psi_{0}(\alpha)| \hat{\mathcal{H}}_{\rm MF}(\alpha)|\psi_{0}(\alpha)\rangle/N$, the appearance of a finite order parameter $\tilde{\alpha}$ corresponds to the instability of the normal ground state, ${\cal E}(\alpha)\leq{\cal E}(0)$. Since we are interested only in the onset of the phase transition, in this derivation we retain only terms up to second order in $\alpha$.

First of all, we remind the reader that  $\bar{\epsilon}_n(0)$ and $\ket{\psi_n(0)}$ are the eigenvalues and eigenvectors of the operator $\hat{\mathcal{H}}_0\equiv{\hat{\mathcal{H}}_{\rm e}}/{N}$ (see Eq.~(8) in the main text). 
  
Now we expand the energy functional $E[\alpha,\psi]$ up to the second order in $\alpha$. Hence, we can consistently approximate the diamagnetic magnetization with the one in absence of light, $\mel{\psi}{\hat{\mathcal{M}}_{\rm d} (0)}{\psi}\approx  \mel{\psi_0(0)}{\hat{\mathcal{M}}_{\rm d} (0)}{\psi_0(0)}= \mathcal{M}^{0,0}_{\rm d}(0)$. Up to the second order in $\alpha$ the energy functional reads
\begin{equation}
    \label{eq: Energy functional}
  E[\alpha,\psi] =\Omega\alpha^2 + \mel{\psi}{\hat{\mathcal{H}}_0}{\psi} + \mel{\psi}{\hat{\mathcal{M}}_{\rm p} (0)}{\psi} \alpha~,
\end{equation}
where $\Omega\equiv \omega_{\rm c} +  \mathcal{M}^{0,0}_{\rm d}(0) / 2  $. By minimizing with respect to $\psi$ we get ${\cal E}(\alpha)\equiv{\rm min}_{\psi}E[\alpha,\psi]$.
Hence, we need to calculate
\begin{equation}
    \label{eq: Energy functional 2}
 {\cal E}(\alpha) =\Omega\alpha^2+{\rm min}_{\psi}\{\mel{\psi}{(\hat{\mathcal{H}}_0 + \hat{\mathcal{M}}_{\rm p} (0) \alpha)}{\psi} \}~,
\end{equation}
Using second-order perturbation theory we obtain
\begin{eqnarray}
\label{eq:HAlt2}
&~&{\rm min}_{\psi}\{\mel{\psi}{(\hat{\mathcal{H}}_0 + \hat{\mathcal{M}}_{\rm p} (0) \alpha)}{\psi} \}=\\ &=& \bar{\epsilon}_0(0)+\frac{\chi_{\rm p}(0)}{2}\alpha^2+ { \mathcal{M}^{0,0}_{\rm p}(0)}\alpha~,\nonumber
\end{eqnarray}
where
\begin{eqnarray}
\label{eq: M_p0 definition}
{ \mathcal{M}^{0,0}_{\rm p}(0)}&\equiv& \mel{\psi_0}{\hat{\mathcal{M}}_{\rm p} (0)}{\psi_0}~,\\
\label{eq: chi_p0 definition}
{\chi_{\rm p}} (0) &\equiv& -2 \sum_{n\neq 0} \frac{\abs{\mel{\psi_n}{\hat{\mathcal{M}}_{\rm p} (0)}{\psi_0}}^2}{\bar{\epsilon}_n(0) - \bar{\epsilon}_0(0)} \leq 0~.
\end{eqnarray}

At this point, there are two different cases to distinguish: {\it i)} the normal ground state paramagnetic moment is zero,  ${ \mathcal{M}^{0,0}_{\rm p}(0)}=0$ or {\it ii)} the normal phase has a finite paramagnetic character ${ \mathcal{M}^{0,0}_{\rm p}(0)}\neq0$.
In the case {\it i)}, Eq.~\eqref{eq:HAlt2} reads
\begin{equation}
 \label{eq:HAlt3A}
 {\cal E}(\alpha)= \bar{\epsilon}_0(0)+\big(\Omega +\frac{\chi_{\rm p}(0)}{2}\big)\alpha^2~.
\end{equation} 

It is possible to define the magnetic susceptibility $\chi_{\rm M}$ as the concavity of the molecular energy $\bar{\epsilon}_0(\alpha)$

\begin{equation}
 \label{eq:HAltA}
     \chi_{\rm M} \equiv -\frac{1}{2} \partial^2_{\alpha}  \bar{\epsilon}_0(\alpha) \Big|_{{\alpha}=0}~.
\end{equation}
In order to calculate $ \chi_{\rm M}$ it is useful to notice that $ \bar{\epsilon}_0(\alpha)= {\cal E}(\alpha)-\omega_{\rm c}\alpha^2~$.
By comparing Eq.~\eqref{eq:HAlt3A} with Eq.~\eqref{eq:HAltA}, we enstablish the a relation between the magnetic susceptibility $\chi_{\rm M}$ and the following microscopic quantities
\begin{eqnarray}
\label{eq: chi_M definition}
  - \chi_{\rm M}+\omega_{\rm c} &=&\big(\Omega +\frac{\chi_{\rm p}(0)}{2}\big)~,
\end{eqnarray} 
or equivalently $- \chi_{\rm M}={ \mathcal{M}^{0,0}_{\rm d}(0)}/2+\chi_{\rm p}(0)/{2}$.
Hence, the criterion for the occurrence of photon condensation ${\cal E}(\alpha)\leq{\cal E}(0)$ is given by
\begin{equation}
\label{eq: chi_M criteria for superradiance to occur}
   \chi_{\rm M}\geq  \omega_{\rm c}~.
\end{equation} 

Now, we focus on the case {\it ii)}, where ${ \mathcal{M}^{0,0}_{\rm p}(0)}\neq 0$.
Eq.\eqref{eq: Energy functional 2} can be expressed at the linear order in $\alpha$ as
\begin{equation}
    \label{eq:H9}
  {\cal E}(\alpha) =  \bar{\epsilon}_0(0)+ { \mathcal{M}^{0,0}_{\rm p}(0)}\alpha~.
\end{equation}
In this case, it is always possible to find a $\bar{\alpha}$ such that ${\cal E}(\bar{\alpha})\leq{\cal E}(0)$ by choosing $\bar{\alpha}$ to have a different sign with respect to 
${ \mathcal{M}^{0,0}_{\rm p}(0)}$. In this case, the system is always unstable and displays photon condensation.
$\mathcal{M}^{0,0}_{\rm p}(0)$ can be different from zero if an additional classic field is introduced. Fig.~1(b) reproduces both conditions \eqref{eq: chi_M criteria for superradiance to occur} and \eqref{eq:H9} in the case of an ensemble of squared rings, varying one hopping parameter and the external classic field.
\section{\label{subsec: sm Van Vleck}Microscopic theory of Van Vleck paramagnetism}

In this Section, we briefly remind the reader about the microscopic theory of Van Vleck paramagnetism. 

We consider a generic many-electron Hamiltonian of the form
\begin{equation}
\hat{\cal{H}}_0=\sum_{k}\Big[\frac{\hat{\bm{p}}^2_k}{2m}+V(\hat{\bm{r}}_k)\Big] +\frac{1}{2}\sum_{k\neq k^\prime} v(\hat{\bm{r}}_k-\hat{\bm{r}}_k^\prime)~,
\end{equation}
where $m$ is the electronic mass, $\hat{\bm{p}}_k$ is the momentum of the $k-$ electron, $V(r)$ is a external potential and $v(\hat{\bm{r}}_k-\hat{\bm{r}}_k^\prime)$ is the electron-electron interaction.
The coupling to a uniform magnetic field ${\bm B}$ is made as customary via the minimal coupling substitution, i.e.~$\hat{\bm{p}}_k\to \hat{\bm{p}}_k+(e/c)\bm{A}(\bm{r}_k)$.

In the symmetric gauge, the vector potential can be expressed as follows
\begin{equation}\label{eq:symmetric gauge static B field}
\bm{A}(\bm{r})=\frac{1}{2}B\hat{\bm{e}}_z\times \bm{r},
\end{equation}
where $\bm{B}=B {\bm e}_z$ is the magnetic field. Notice that the symmetric-gauge vector potential obeys also the Coulomb condition, i.e. $\bm{\nabla}\cdot\bm{A}(\bm{r})=0$. 
Carrying out the minimal coupling in the symmetric gauge, we find the following Hamiltonian of the many-electron systems coupled to a uniform external magnetic field $\bm{B}$:
\begin{equation}
\hat{\cal H}=\hat{\cal H}_0+\frac{e}{2mc} \bm{B}\cdot \hat{\bm{L}} +\frac{e^2}{8mc^2}B^2\sum_{k=1}^N |\hat{\bm{r}}_k|^2~,
\end{equation}
where $\hat{\bm{L}} =\sum_{k=1}^N(\hat{\bm{r}}_k\times \hat{\bm{p}}_k)$ is the (paramagnetic) angular momentum operator and the last term is the magnetostatic energy. Note that, accordingly to the main text, we have neglected the Zeeman coupling.

The magnetization $\hat{\bm{M}}$ is related to the angular momentum and magnetic field by the following relation:
\begin{align}
\hat{\bm{M}}&\equiv -\frac{ei}{2c}   \sum_{k=1}^N\hat{\bm r}_k  \times  [\mathcal{H}, \hat{\bm r}_k ]\nonumber\\
&= -\frac{e}{2mc}\hat{\bm{L}}- \frac{e^2}{2mc^2} \sum_{k=1}^N  \hat{\bm{r}}_k\times \bm{A}(\bm{r}_k)\nonumber\\
&=- \frac{e}{2mc}\hat{\bm{L}}- \frac{e^2B}{4mc^2}\Big( \sum_{k=1}^N  |\hat{\bm{r}}_k|^2\Big) \hat{\bm{z}}\nonumber\\
&\equiv \hat{\bm{M}}_{\rm p}+\hat{\bm{M}}_{\rm d}~,
\end{align}
where the first (second) term is the paramagnetic (diamagnetic) contribution.

In order to find the orbital magnetic susceptibility, we need to study the energy variation under the applied magnetic field. This splits into two terms:
$\Delta E=\Delta E_{\rm P}
+\Delta E_{\rm D}$. Introducing the exact eigenstates and eigenvalues of the Hamiltonian $\hat{\cal H}_0$, we find:
\begin{align}
\Delta E_{\rm D}&\equiv\frac{e^2}{8mc^2}B^2\langle \psi_0|\sum_{k=1}^N| \hat{\bm{r}}_k|^2|\psi_0\rangle\nonumber\\
&=-\frac{1}{2} \langle\psi_0|\hat{{M}}_{{\rm d},z}|\psi_0\rangle B^2~,
\end{align}
where $\hat{{M}}_{{\rm d},z}$ is the $\bm{e}_z$ component of the vector $\hat{\bm{M}}_{\rm d}$ and
\begin{align}
\Delta E_{\rm P}&\equiv\Big(\frac{e}{2mc}\Big)^2\sum_{n\neq 0}\frac{|\langle \psi_n|  \bm{B}\cdot \hat{\bm{L}} |\psi_0\rangle|^2}{E_n-E_0}\nonumber\\
&= \sum_{n\neq 0}\frac{|\langle \psi_n|  \bm{B}\cdot \hat{\bm{M}}_{\rm p} |\psi_0\rangle|^2}{E_n-E_0}~.
\end{align}
$\Delta E_{\rm B}$ is the magnetic contribution, while
by construction $\Delta  E_{\rm D}>0$ and $\Delta  E_{\rm P}<0$  i.e. they are a diamagnetic and a paramagnetic contribution, respectively. If the first excited state $n=1$ is nearly degenerate with the ground state $E_{1}-E_{0}\approx 0$, the paramagnetic contribution is dominant and the system has a paramagnetic response (known as Van Vleck paramagnetism). 

The paramagnetic contribution can be recast as $\Delta E_{\rm P}=-(1/2){\chi_{\rm p}}B^2$ where $\chi_{\rm p}$ is the magnetization response function,
\begin{eqnarray}
\chi_{{\rm p}}&\equiv&-2\sum_{n\neq 0}\frac{|\langle \psi_n|  \hat{\bm{z}}\cdot \hat{\bm{M}}_{\rm p} |\psi_0\rangle|^2}{E_n-E_0},\\
&=&-2\Big(\frac{e}{2mc}\Big)^2\sum_{n\neq 0}\frac{|\langle \psi_n|  \hat{\bm{z}}\cdot \hat{\bm{L}} |\psi_0\rangle|^2}{E_n-E_0},\label{eq:ChiLL}
\end{eqnarray}

Thus, orbital paramagnetism is governed by the angular momentum-angular momentum response function. From the expression for the response function in Eq. \eqref{eq:ChiLL}, it can be seen that non-zero orbital paramagnetic response occurs only if the system does not have rotational invariance around the $\hat{\bm{z}}$ axis. Indeed, if $[\hat{\cal H}_0, \hat{\bm{z}}\cdot \hat{\bm L}]=0$, one can choose a common eigenstate basis of the energy and the angular momentum projection $\hat{\bm{z}}\cdot {\bm L}$ and the matrix element $\bra{\psi_0} \hat{\bm{z}}\cdot \hat{\bm L}\ket{\psi_n}$ vanishes, precluding the presence of Van Vleck paramagnetism in rotationally invariant systems, such as closed shell atoms.

\section{\label{subsec: sm Bosonization and polaritons study}Bosonization and polaritons}
In order to study the polaritonic properties of our coupled light-matter system, we start from Eq.~(4). First of all, as stated in the main text, we shift the cavity photon operators, $\aop =\alpha \sqrt{N}+\delta \aop $, where $\delta \aop$ describes zero-average fluctuations around the mean-field solution $\alpha\sqrt{N}$. Hence, in the thermodynamic limit, and expanding the shifted Hamiltonian up to second order in the fluctuations $\delta \aop$, we find the following Hamiltonian:
\begin{eqnarray}
    \hat{\mathcal{H}} &=& \omega_{\rm c} \left[ \delta\adop \delta\aop + \alpha \sqrt{N} \left( \delta\aop + \delta\adop \right) + \alpha^2 N \right] \\
    &+& \sum_{k = 1}^N \hat{ h}_{{\rm e},k} (\alpha)
    + \frac{\sqrt{N}}{2 } \hat{\mathcal{M}}_{\rm p} (\alpha) \left(\delta \aop +\delta \adop \right) \nonumber\\ &+& \frac{1}{8 } \hat{\mathcal{M}}_{\rm d} (\alpha) \left(\delta \aop + \delta\adop \right)^2~. \nonumber
\end{eqnarray}

It is now useful to introduce fermionic operators, $\hat{\varphi}_{l, k}$ and $\hat{\varphi}_{l, k}^\dagger$, that  destroy and create an electron in the 
energy eigenstate $\ket{\varphi_{l}(\alpha) }_k$ of the Hamiltonian $\hat{h}_{{\rm e},k}(\alpha)$, defined in Eq.~(6) of the main-text, i.e.~$\hat{h}_{{\rm e},k}(\alpha) \ket{\varphi_{l}(\alpha) }_k = \epsilon_l (\alpha) \ket{\varphi_{l}(\alpha) }_k$.
Notice that $\hat{\varphi}_{l, k}$ depends on the photonic mean-field $\alpha$ and we dropped that explicit dependence for the ease of readability.
Here, the eigenvalues $\epsilon_l (\alpha)$ are $k$-independent, since all molecules are identical. 
Expressing $\hat{\mathcal{H}}$ in terms of the operators $\hat{\varphi}_{l, k}$, we obtain
\begin{eqnarray}\label{eq:sm_Hdeltaa}
\hat{\mathcal{H}} &=& \omega_{\rm c} \left[\delta \adop\delta \aop + \alpha \sqrt{N} \left(\delta \aop + \delta\adop \right) + \alpha^2 N \right] \\
&+& \sum_{l = 0}^{n_{\rm s} - 1} \epsilon_l(\alpha) \sum_{k = 1}^N \hat{\varphi}_{l, k}^\dagger\hat{\varphi}_{l, k} \nonumber \\
&+& \frac{1}{2 \sqrt{N}} \left( \delta\aop + \delta\adop \right) \sum_{l, m = 0}^{n_{\rm s} - 1}\mathcal{M}_{\rm p}^{l, m} (\alpha) \sum_{k = 1}^{N} \hat{\varphi}_{l, k}^\dagger\hat{\varphi}_{m, k} \nonumber \\
&+& \frac{1}{8N } \left(\delta \aop + \delta\adop \right)^2 \sum_{l, m = 0}^{n_{\rm s} - 1}\mathcal{M}_{\rm d}^{l, m} (\alpha)  \sum_{k = 1}^{N} \hat{\varphi}_{l, k}^\dagger\hat{\varphi}_{m, k}~,\nonumber 
\end{eqnarray}
where
\begin{align}
\mathcal{M}_{\rm p}^{l, m} (\alpha) &= \sum^N_{k=1} \tensor[_{k}]{\langle}{} {\varphi_{l}(\alpha)}|{\hat{\cal M}_{\rm p}(\alpha)}\ket{\varphi_{m}(\alpha)}_k\label{eq:sm_Mp}\nonumber\\
&=- \frac{2 B_0}{\sqrt{N}}  \sum^N_{k=1} \tensor[_{k}]{\langle}{}{\varphi_{l}(\alpha)}|{\hat{M}_{z,k}(\alpha)}\ket{\varphi_{m}(\alpha)}_k 
\end{align}
and
\begin{align}
\mathcal{M}_{\rm d}^{l, m} (\alpha) &= \sum^N_{k=1} \tensor[_{k}]{\langle}{} {\varphi_{l}(\alpha)}|{\hat{\cal M}_{\rm d}(\alpha)}\ket{\varphi_{m}(\alpha)}_k\label{eq:sm_Md}\nonumber\\
&=- \frac{2 B_0}{\sqrt{N}}  \sum^N_{k=1} \tensor[_{k}]{\langle}{} {\varphi_{l}(\alpha)}|{\partial_\alpha \hat{M}_{z,k}(\alpha)}\ket{\varphi_{m}(\alpha)}_k~,
\end{align}
In writing Eq.~\eqref{eq:sm_Hdeltaa} we used that the matrix elements $\tensor[_{k}]{\langle}{}{\varphi_{l}(\alpha)}|{\hat{M}_{z,k}(\alpha)}\ket{\varphi_{m}(\alpha)}_k$ and $ \tensor[_{k}]{\langle}{}{\varphi_{l}(\alpha)}|{\partial_\alpha \hat{M}_{z,k}(\alpha)}\ket{\varphi_{m}(\alpha)}_k$ are independent of the $k$ index.
Making use of the collective notation 
\begin{equation}\label{eq:Sigma_operators}
\hat{\Sigma}_{l, m} \equiv \sum_k \hat{\varphi}_{l, k}^\dagger\hat{\varphi}_{m, k}~,
\end{equation}
 the Hamiltonian takes a more compact form
\begin{eqnarray}
\label{eq: alpha-shifted second order hamiltonian in the energy basis (collective)}
    \hat{\mathcal{H}} &=& \omega_{\rm c} \left[ \delta \adop \delta \aop + \alpha \sqrt{N} \left( \delta\aop + \delta \adop \right) + \alpha^2 N \right]   \\
    &+&  \sum_{l = 0}^{n_{\rm s} - 1} \epsilon_l (\alpha) \hat{\Sigma}_{l, l}+\frac{1}{2 \sqrt{N}} \left( \aop + \adop \right) \sum_{l, m = 0}^{n_{\rm s} -1} \left[ \mathcal{M}_{\rm p}^{l, m} ( \alpha ) \hat{\Sigma}_{l, m} \right] \nonumber \\
    &+& \frac{1}{8N} \left( \aop + \adop \right)^2 \sum_{l, m = 0}^{n_{\rm s} -1} \left[ \mathcal{M}_{\rm d}^{l, m} ( \alpha ) \hat{\Sigma}_{l, m} \right]~.\nonumber
\end{eqnarray}

By following Ref.~\cite{kurucz_pra_2010}, we focus on the symmetric Hilbert subspace, which is spanned  by the occupation number states defined as
\begin{eqnarray}\label{eq:occstate}
&&\ket{n_0,m_1,\cdots,p_{n_{\rm s}-1}}=\frac{1}{\sqrt{n!m!\cdots p!}}
\sum_{\rm perm} 
\ket{\varphi_0(\alpha)}_1\cdots  \nonumber\\
&\times&\ket{\varphi_0(\alpha)}_n \ket{\varphi_1(\alpha)}_{n+1}\cdots \ket{\varphi_1(\alpha)}_{n+m}\dots
\nonumber\\
&\times&\ket{\varphi_{n_{\rm s}-1}(\alpha)}_{N-p+1}\cdots \ket{\varphi_{n_{\rm s}-1}(\alpha)}_{N}~, 
\end{eqnarray}
which means that $n$ molecules have a single electron that occupies the state $\ket{\varphi_0(\alpha)}$,
$m$ molecules have a single electron that occupies the state $\ket{\varphi_1(\alpha)}$, $\ldots$, and 
$p$ molecules have a single electron that occupies the state   $\ket{\varphi_{n_{\rm s}-1}(\alpha)}$, such that $N=n+m+\ldots+p$.
By applying an occupation number state on collective operators defined in Eq.~\eqref{eq:Sigma_operators}, we find the following properties
\begin{eqnarray}
\hat{\Sigma}_{l,l'}\ket{n_l,m_{l'},\ldots}&=&\sqrt{(n+1)m} \nonumber\\
&\times&\ket{(n+1)_l,m_{l'}-1,\ldots}, \label{eq:Sigma_prop1} \\
\hat{\Sigma}_{l,l}\ket{n_l,m_{l'},\ldots}&=&n\ket{n_l,m_{l'},\ldots}, \label{eq:Sigma_prop2}
\end{eqnarray}
where $l \neq l'$. Moreover, the collective operators fulfill the commutator identity $[\hat{\Sigma}_{l,l'}, \hat{\Sigma}_{m,m'}]=\delta_{l',m}\hat{\Sigma}_{l,m'}-
\delta_{m',l}\hat{\Sigma}_{m,l'}$. 
Here, we introduce $n_{\rm s}-1$ couple of bosonic creation and annihilation operators 
 $\hat{b}_l^\dagger$ and $\hat{b}_l$ such that
\begin{eqnarray}
\hat{b}_l \ket{n_0,m_l,\cdots}&=&\sqrt{m}
 \ket{(n+1)_0,(m-1)_l,\cdots}~, \label{eq:b_prop1}\\
 \hat{b}_l^\dagger \hat{b}_l \ket{n_0,m_l,\cdots}&=&m
\ket{n_0,m_l,\cdots}~,  \label{eq:b_prop2}
\end{eqnarray}
where $l=1,\ldots,n_{\rm s}-1$, and the mean-field groundstate $\ket{\psi_0(\alpha)}=\ket{N_0,0,\cdots,0}=\prod^N_{k=1} \ket{\varphi_0(\alpha)}_k $,  that corresponds to all molecules with a single electron in the state $\ket{\varphi_0(\alpha)} $,  acts as the vacuum state for any $\hat{b}_l$. 
Creation operators $\hat{b}^\dagger_l$ applied on the vacuum state $\ket{\psi_0(\alpha)}$ describe the collective matter excitations, which represent the bright modes.
By comparing Eqs.~\eqref{eq:Sigma_prop1}-\eqref{eq:Sigma_prop2} with Eqs.~\eqref{eq:b_prop1}-\eqref{eq:b_prop2}, we write the collective operators
 in terms of the bosonic fields accordingly to a multilevel Holstein-Primakoff transformation~\cite{kurucz_pra_2010}, 
\begin{eqnarray}
\hat{\Sigma}_{0,0}&=&N-\sum_{l>0} \hat{b}_l^\dagger\hat{b}_l~,\\
\hat{\Sigma}_{l,0}&=&\hat{b}_l^\dagger \sqrt{N-\sum_{l^\prime>0} \hat{b}_{l^\prime}^\dagger\hat{b}_{l^\prime} }~,\\
\hat{\Sigma}_{l,l^\prime}&=&\hat{b}_l^\dagger\hat{b}_{l^\prime}~,
\end{eqnarray}
where $l, l^\prime>0$.
In the proximity of the mean-field matter ground state $\ket{\psi_0(\alpha)}$, namely for a small number of collective matter excitations, one can approximate
\begin{eqnarray}
\hat{\Sigma}_{l,0}&\approx &\sqrt{N}\hat{b}_l^\dagger~.
\end{eqnarray}
By using the matter collective bosonic fields  $\hat{b}_l^\dagger$ and $\hat{b}_l$ to rewrite the Hamiltonian in Eq.~\eqref{eq: alpha-shifted second order hamiltonian in the energy basis (collective)}, and we get rid of all terms beyond the second order in the (light and matter) bosonic fields 
\begin{eqnarray}\label{eq:Hquadratic}
    \hat{\mathcal{H}} &\simeq& \omega_{\rm c} \delta\adop \delta\aop + \omega_{\rm c} \sqrt{N} \left[ \alpha + \frac{\mathcal{M}_{\rm p}^{0,0} ( \alpha )}{2 \omega_{\rm c}} \right] \left(\delta \aop + \delta\adop \right)  \nonumber \\
    &+&N \omega_{\rm c} \alpha^2 + N \epsilon_0 (\alpha) + \sum_{l = 1}^{n_{\rm s} - 1} \left[ \epsilon_l (\alpha) - \epsilon_0 (\alpha) \right] \hat{b}_l^\dagger \hat{b}_l \nonumber \\
    &+& \frac{1}{2} \left( \delta \aop + \delta \adop \right) \sum_{l = 1}^{n_{\rm s} - 1} \left[ \mathcal{M}_{\rm p}^{l,0} ( \alpha ) \hat{b}_l^\dagger + \mathcal{M}_{\rm p}^{0,l} ( \alpha ) \hat{b}_l \right] \nonumber\\
    &+& \frac{1}{8} \left( \delta  \aop + \delta  \adop \right)^2 \mathcal{M}_{\rm d}^{0, 0} ( \alpha )~,
\end{eqnarray}

This Hamiltonian still depends upon the mean-field displacement $\alpha$.
 In order to fix the value of $\alpha$, we select the value which nullifies the derivative of the energy functional ${\cal E}(\alpha)=\bra{\psi_0(\alpha)} \hat{\cal H}_{\rm MF} (\alpha)\ket{\psi_0(\alpha)}/N$ with respect $\alpha$, i.e.
\begin{equation}
\frac{d {\cal E}(\alpha)}{d \alpha}=2 \omega_{\rm c} \alpha+
\frac{d \bar{\epsilon}_0(\alpha)}{d \alpha}~.
\end{equation}
By using the Hellmann–Feynman theorem to compute $({d  \bar{\epsilon}_0(\alpha)}/{d \alpha})$, 
\begin{eqnarray}
\frac{d\bar{\epsilon}_0(\alpha)}{d \alpha}&=&\frac{1}{N} \sum^N_{k=1} \bra{\psi_0(\alpha)} \frac{d \hat{h}_{{\rm e},k}(\alpha)}{d \alpha}\ket{\psi_0(\alpha)}~, \\
&=&\bra{\psi_0(\alpha)} \hat{\cal M}_{\rm p}(\alpha) \ket{\psi_0(\alpha)}=\mathcal{M}_{\rm p}^{0, 0} (\alpha),\nonumber
\end{eqnarray}
we can write a non-linear equation to determine $\tilde{\alpha}$: 
\begin{equation}\label{eq: sm solutions for alpha}
\frac{d {\cal E}(\alpha)}{d \alpha}\bigg|_{\alpha=\tilde{\alpha}}=2\omega_{\rm c} \tilde{\alpha}+\mathcal{M}_{\rm p}^{0, 0} (\tilde{ \alpha} )=0~.
\end{equation}
By imposing $\alpha=\tilde{\alpha}$, the linear terms in the bosonic operators in Eq.~\eqref{eq:Hquadratic} vanish. We clearly see that Eq.~(\ref{eq: sm solutions for alpha}) coincides with Eq.~(10) in the main text.

Once we set $\alpha=\tilde{\alpha}$, the approximate Hamiltonian in Eq.~\eqref{eq:Hquadratic} becomes a {\it polaritonic} Hamiltonian $\hat{\cal H}_{\rm pol}$ the low-energy hybrid excitations (polaritons) on top of the mean-field ground-state solution, 
\begin{eqnarray}
\label{eq: sm quadratic bosonic Hamiltonian}
\hat{\cal H}_{\rm pol} &=& \omega_{\rm c} \delta \adop \delta \aop 
+ \sum_{l=1}^{n_{\rm s} - 1} [\epsilon_{l}(\tilde{\alpha}) -\epsilon_{0}(\tilde{\alpha})]\hat{b}_l^\dagger \hat{b}_l \nonumber\\
&+& \frac{1}{2} (\delta \aop + \delta \adop) \sum_{l=1}^{n_{\rm s} - 1} \left[ \mathcal{M}_{\rm p}^{l,0} ( \tilde{\alpha} ) \hat{b}_l^\dagger + \mathcal{M}_{\rm p}^{0,l} (\tilde{\alpha}) \hat{b}_l \right] \nonumber \\
&+& \frac{1}{8} \mathcal{M}_{\rm d}^{0, 0} ( \tilde{\alpha} ) (\delta \aop + \delta \adop)^2~.
\end{eqnarray}
%
Polaritons are linear combinations of light and matter operators,
\begin{equation}
\label{eq:polariton}
\hat{p}_{\nu}=X_\nu \delta \hat{a} + Y_\nu \delta \hat{a}^\dagger +
\sum^{n_{\rm s}-1}_{l=1} \big(W_{\nu,l}   \hat{b}_l + Z_{\nu,l} \hat{b}^\dagger_l\big)~.
\end{equation}

Being the polariton a proper bosonic excitation of the system, the operator $\hat{p}_{\nu}$ fulfills the equation of motion of a harmonic ladder operator
\begin{equation}
\label{eq:polaritonEOM}
[\hat{\cal H}_{\rm pol},\hat{p}_{\nu}]= -\Omega_{{\rm p},\nu}  \hat{p}_{\nu}~.
\end{equation}
Being the polariton a combination of $\hat{a},\hat{a}^\dagger,\hat{b}_l,\hat{b}_l^\dagger $, in order to calculate Eq.~\eqref{eq:polaritonEOM} we need the equations of motion for these light and matter operators.
These are given by the following commutators
\begin{eqnarray}
[\hat{\cal H}_{\rm pol},\delta \aop]&=&-\omega_{\rm c} \delta \aop - \frac{1}{4} \mathcal{M}_{\rm d}^{0, 0} (\tilde{\alpha})(\delta\aop+\delta\adop) \nonumber \\
&-& \frac{1}{2} \sum^{n_{\rm s}-1}_{l=1} (\mathcal{M}_{\rm p}^{l,0}(\tilde{\alpha}) \hat{b}_l^\dagger + \mathcal{M}_{\rm p}^{0,l}(\tilde{\alpha}) \hat{b}_l)~,
\end{eqnarray}
\begin{eqnarray}
[\hat{\cal H}_{\rm pol},\delta \adop]&=&\omega_{\rm c}  \delta \adop + \frac{1}{4} \mathcal{M}_{\rm d}^{0, 0}(\tilde{\alpha})(\delta \aop+\delta \adop)\nonumber \\
&+& \frac{1}{2}
 \sum^{n_{\rm s}-1}_{l=1} (\mathcal{M}_{\rm p}^{l,0} (\tilde{\alpha}) \hat{b}_l^\dagger + \mathcal{M}_{\rm p}^{0,l}(\tilde{\alpha}) \hat{b}_l)~,
\end{eqnarray}
\begin{equation}
[\hat{\cal H}_{\rm pol},\hat{b}_l]=-[\epsilon_{l}(\tilde{\alpha}) -\epsilon_{0}(\tilde{\alpha})] \hat{b}_l -
  \frac{1}{2} \mathcal{M}_{\rm p}^{0,l} (\tilde{\alpha})  (\delta\aop+\delta\adop)~,
\end{equation}
\begin{equation}
[\hat{\cal H}_{\rm pol},\hat{b}_l^\dagger]=\-[\epsilon_{l}(\tilde{\alpha}) -\epsilon_{0}(\tilde{\alpha})] \hat{b}_l^\dagger +
 \frac{1}{2} \mathcal{M}_{\rm p}^{l,0}(\tilde{\alpha}) (\delta\aop+\delta\adop)~,
\end{equation}
By using these commutators and Eq.~\eqref{eq:polariton}, the equation of motion expressed in Eq.~\eqref{eq:polaritonEOM} can be mapped in the following eigenvalue problem
\begin{equation}
\label{eq:eigenvalueP}
\Xi {\bm v}_{\nu}=\Omega_{\rm p,\nu}   {\bm v}_{\nu}~, 
\end{equation}
where ${\bm v}_\nu=(X_\nu,Y_\nu,{\bm W}_\nu,{\bm Z}_\nu)^{\rm T}$, and
\begin{equation}
\label{eq:sm_Hopfield_Matrix}
\Xi=
\begin{pmatrix}
\omega_{\rm c} + \frac{1}{4} \mathcal{M}_{\rm d}^{0,0}(\tilde{\alpha}) & - \frac{1}{4} \mathcal{M}_{\rm d}^{0,0}(\tilde{\alpha}) &   {\bm g}(\tilde{\alpha}) & - {\bm g}^*(\tilde{\alpha})\\
\frac{1}{4} \mathcal{M}_{\rm d}^{0,0}(\tilde{\alpha}) & -\omega_{\rm c} - \frac{1}{4} \mathcal{M}_{\rm d}^{0,0}(\tilde{\alpha})&    {\bm g}(\tilde{\alpha}) & -  {\bm g}^*(\tilde{\alpha})\\
  {{\bm g}^*}^{\rm T}(\tilde{\alpha}) &-{{\bm g}^*}^{\rm T}(\tilde{\alpha}) & {\bm \Omega}(\tilde{\alpha}) & 0\\
  {\bm g}^{\rm T}(\tilde{\alpha}) &-{\bm g}^{\rm T}(\tilde{\alpha}) & 0 & - {\bm \Omega}(\tilde{\alpha})\\
\end{pmatrix}~,
\end{equation}
where ${\bm g}(\tilde{\alpha})= ({1}/{2}) [\mathcal{M}_{\rm p}^{1,0}(\tilde{\alpha}), \dots ,\mathcal{M}_{\rm p}^{n_{\rm s}-1,0}(\tilde{\alpha})]$ and $ {\bm \Omega}(\tilde{\alpha}) = {\rm diag} [\epsilon_{1}(\tilde{\alpha}) -\epsilon_{0}(\tilde{\alpha}), \dots, \epsilon_{n_{\rm s}-1}(\tilde{\alpha}) -\epsilon_{0}(\tilde{\alpha})]$.

Eigenvalues of the Hopfield matrix $\Xi$ can be determined by the roots of the determinant ${\cal D}(\Omega_{{\rm p},\nu})$
\begin{equation}
\label{eq:det}
{\cal D}(\Omega_{{\rm p},\nu})={\rm Det}(\Omega_{{\rm p},\nu} \openone-\Xi)=0~.   
\end{equation}
The determinant ${\cal D}(\Omega_{{\rm p},\nu})$ can be calculated by using the following algebraic property of the block matrices
\begin{equation*}
{\rm Det}
\begin{pmatrix}
A&B\\
C&D
\end{pmatrix}
={\rm Det}(D){\rm Det}(A-BD^{-1}C)~,
\end{equation*}
which leads to
\begin{widetext}
\begin{eqnarray}
{\cal D}(\Omega_{{\rm p},\nu})&=&
\prod^{n_{\rm s}-1}_{l=1}\{\Omega_{{\rm p},\nu}^2 - [\epsilon_{l}(\tilde{\alpha}) -\epsilon_{0}(\tilde{\alpha})]^2\}\\
&\times&{\rm Det}
\begin{pmatrix}
\omega_{\rm c} + [\frac{1}{4} \mathcal{M}_{\rm d}^{0, 0} (\tilde{\alpha})+ \frac{1}{4} \chi^{(\tilde{\alpha})}_{\rm p}(\Omega_{{\rm p},\nu})]
-\Omega_{{\rm p},\nu}&- [\frac{1}{4} \mathcal{M}_{\rm d}^{0, 0}(\tilde{\alpha}) + \frac{1}{4} \chi^{(\tilde{\alpha})}_{\rm p}(\Omega_{{\rm p},\nu})]  \\
 [\frac{1}{4} \mathcal{M}_{\rm d}^{0, 0}(\tilde{\alpha}) + \frac{1}{4} \chi^{(\tilde{\alpha})}_{\rm p}(\Omega_{{\rm p},\nu})]& - \omega_{\rm c} -  [\frac{1}{4} \mathcal{M}_{\rm d}^{0, 0} (\tilde{\alpha})+ \frac{1}{4} \chi^{(\tilde{\alpha})}_{\rm p}(\Omega_{{\rm p},\nu})]-\Omega_{{\rm p},\nu}
\end{pmatrix}~, \nonumber
\end{eqnarray}
\end{widetext}
where
\begin{eqnarray}
\label{eq: sm chi di omega}
\chi^{(\tilde{\alpha})}_{\rm p}(\omega)&=&
2 \sum^{n_{\rm s}-1}_{l=1} 
\frac{\abs{\mathcal{M}_{\rm p}^{l,0}(\tilde{\alpha})}^2 [\epsilon_l(\tilde{\alpha})-\epsilon_0(\tilde{\alpha})]}{\omega^2-[\epsilon_l(\tilde{\alpha})-\epsilon_0(\tilde{\alpha})]^2}~.
\end{eqnarray}
The static limit  of $\chi^{(\tilde{\alpha})}_{\rm p}(\omega)$ generalizes the static paramagnetic susceptibility 
for a finite value of $\tilde{\alpha}$, such that $\chi^{(0)}_{\rm p}(0)=\chi_{\rm p}(0)$, where $\chi_{\rm p}(0)$ has been defined in the main text as
%
\begin{equation}
\label{eq:chipS}
\chi_{\rm p}(0)= -2 \sum_{n \neq 0}\frac{ \abs{\mel{\psi_n(0)}{\hat{\mathcal{M}}_{\rm p} (0)}{\psi_0(0)}}^2}{\bar{\epsilon}_n(0) - \bar{\epsilon}_0(0)}~.
\end{equation}
To verify this generalization, we note that the matrix element $\mel{\psi_n(0)}{\hat{\mathcal{M}}_{\rm p} (0)}{\psi_0(0)}$ connects the mean-field many-body ground state $\ket{\psi_0(0)}$ with a mean-field many-body state $\ket{\psi_n(0)}$ with only an excited molecule, namely the $N$-tuple $\{l_{1,n},\ldots,l_{k,n},\ldots,l_{N,n}\}$ associated with the many-body excited state $\ket{\psi_n(0)}$ consists of $N-1$ zeroes and a single non-zero element $l_{k^\star,n}=l^\star_n>0$.

Thus, the energy difference between the many-body excited state and the many-body ground state can be expressed in terms of the single-particle energy difference between excited states and the ground state as $\bar{\epsilon}_n(0)-\bar{\epsilon}_0(0)=[\epsilon_{l^\star_n}(0)-\epsilon_0(0)]/N$.
Moreover, since we are dealing with $N$ identical molecules, there are $N$ different excited states (with a single excitation) associated with the gap $\bar{\epsilon}_n(0)-\bar{\epsilon}_0(0)$. 
%
Thanks to these remarks and by means of Eq.~(\ref{eq:sm_Mp}), the static paramagnetic susceptibility (in Eq.~\eqref{eq:chipS}) can be expressed as 
\begin{eqnarray}
\label{eq:chi1s}
\chi_{\rm p}(0)=-2 \sum^{n_{\rm s}-1}_{l=1}\frac{\abs{\mathcal{M}_{\rm p}^{l,0}(0)}^2}{[\epsilon_l({0})-\epsilon_0(0)]}~.
\end{eqnarray}
Comparing this expression with Eq.~\eqref{eq: sm chi di omega} evaluated at $\omega=0$, we conclude that
\begin{eqnarray}
\label{eq:chi=chi}
\chi_{\rm p}^{(0)}(0)=\chi_{\rm p}(0)~.
\end{eqnarray}

With few algebraic manipulations, the determinant can be expressed as
\begin{eqnarray}
{\cal D}(\Omega_{{\rm p},\nu}) &=&
\prod^{n_{\rm s}-1}_{l=1}\{\Omega_{{\rm p},\nu}^2 - [\epsilon_{l}(\tilde{\alpha}) -\epsilon_{0}(\tilde{\alpha})]^2\}\bigg\{\Omega_{{\rm p},\nu}^2 + \nonumber\\ &-& \omega_{\rm c}\left[ \omega_{\rm c} 
+  \left( \frac{1}{2} \mathcal{M}_{\rm d}^{0, 0} (\tilde{\alpha})+ \frac{1}{2} \chi^{(\tilde{\alpha})}_{\rm p}(\Omega_{{\rm p},\nu}) \right) \right]\bigg\}~.\nonumber\\
\end{eqnarray}
Hence, polaritonic eigenenergies are given by the following non-linear equation
\begin{equation}\label{eq:eiP_pol}
\Omega_{{\rm p},\nu}^2-\omega_{\rm c}\left\{\omega_{\rm c} + \left[ \frac{1}{2} \mathcal{M}_{\rm d}^{0, 0}(\tilde{\alpha}) + \frac{1}{2} \chi^{(\tilde{\alpha})}_{\rm p}(\Omega_{{\rm p},\nu}) \right]\right\}=0~, 
\end{equation}
where we remind that $\chi^{(\tilde{\alpha})}_{\rm p}(\Omega_{{\rm p},\nu})$ is a function of $\tilde{\alpha}$.
The onset of the superradiant phase transition corresponds to a softening of a polariton, i.e. a polariton with zero energy.
Eq.~\eqref{eq:eiP_pol} for $\Omega_{{\rm p},\nu}=0$ reduces to 
\begin{equation}\label{eq:pol-instability}
\omega_{\rm c} +  \frac{1}{2} \mathcal{M}_{\rm d}^{0, 0}(\tilde{\alpha}) + \frac{1}{2} \chi^{(\tilde{\alpha})}_{\rm p}(0)=0~.  
\end{equation}
We note that for $\tilde{\alpha}=0$, employing Eq.~\eqref{eq:chi=chi}, the previous equation coincides with the instability criterion shown in Eq.~\eqref{eq: chi_M criteria for superradiance to occur}.

\begin{figure}[b]
\centering
\includegraphics[width=\linewidth]{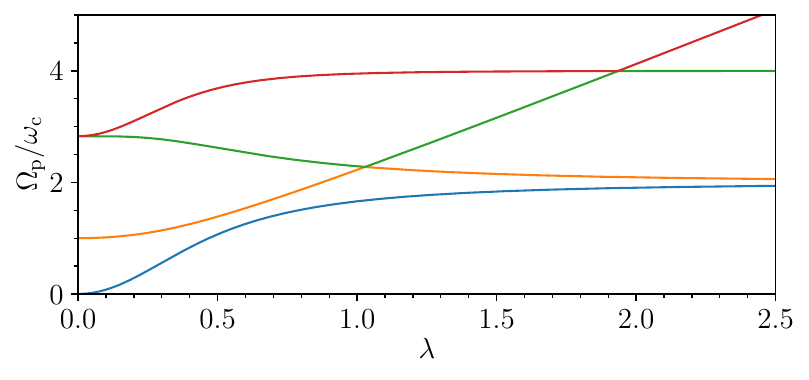}
\caption{(Color online)  Polariton softening at $\lambda \to 0$. Results in this figure have been obtained by setting with $\Theta=0$, $\tau =- \omega_{\rm c}$, and $t=\omega_{\rm c}$. The four polariton energies $\Omega_{\rm p}$ are plotted as functions of $\lambda$. Softening of the lowest polariton mode occurs for $\lambda \to 0$. \label{fig:polaritons_2}}
\end{figure}

In Fig.~\ref{fig:polaritons_2} we show the polariton eigenenergies, by diagonalizing the Hopfield matrix expressed in Eq.~\eqref{eq:sm_Hopfield_Matrix}. In this figure, we illustrate the same quantities as in the analog figure of the main text but for a larger value of $\tau$, i.e.~for $\tau=-\omega_{\rm c}$. In this case, polariton softening occurs in the weak-coupling $\lambda\to 0$ limit. The reason is easy to understand. The criterion (\ref{eq: chi_M criteria for superradiance to occur}) for the occurrence of photon condensation in molecular systems we derived in this Letter depends on the {\it intrinsic} orbital magnetic response $\chi_{\rm M}$ of the molecular system, i.e.~on the dependence of $\chi_{\rm M}$ on the microscopy of the molecular Hamiltonian $\hat{h}_{{\rm e},k}$ in the absence of light-matter interactions. For $\tau=-\omega_{\rm c}$, the susceptibility $\chi_{\rm p}(0)$ diverges due to a zero in the denominator of a term in the sum given in Eq.~\eqref{eq: chi_p0 definition}. This divergence arises from the double degeneracy of the ground state of the molecular system in the absence of light-matter interactions, which results in $\bar{\epsilon}_1-\bar{\epsilon}_0\approx 0$. Due to the explicit structure of $ \chi_{\rm M}$ (see Eq.~\eqref{eq: chi_M definition}), an infinitesimal $\lambda$ is sufficient to achieve a large $\chi_{\rm M}$ that satisfies the criterion (\ref{eq: chi_M criteria for superradiance to occur}).

\end{document}